\documentstyle[psfig]{espart}
\newcounter{refcaption}
\newcounter{nofigs}
\begin{document}
\begin{frontmatter}

\title{{\small J. Phys. I France, Vol. 5, p. 1431-1455
(1995)}\\[1.5cm]
Molecular dynamics of arbitrarily shaped granular particles}

\author{Thorsten P\"oschel\thanksref{bylinetp}}

\address{Arbeitsgruppe Nichtlineare Dynamik, Universit\"at Potsdam, Am
  Neuen\\ Palais, D--14415 Potsdam, Germany.}

\author{Volkhard Buchholtz\thanksref{bylinevb}}

\address{Humboldt--Universit\"at zu Berlin, Institut f\"ur
  Physik, Unter den Linden 6,\\ D--10099 Berlin, Germany}

\thanks[bylinetp]{E-mail: thorsten@hlrsun.hlrz.kfa--juelich.de}
\thanks[bylinevb]{E-mail: volkhard@itp02.physik.hu--berlin.de}
\date{16 July 1995}

\maketitle

\begin{abstract}
  We propose a new model for the description of complex granular
  particles and their interaction in molecular dynamics simulations of
  granular material in two dimensions. The grains are composed of
  triangles which are connected by deformable beams. Particles are
  allowed to be convex or concave. We present first results of
  simulations using this particle model.
\end{abstract}
\end{frontmatter}

\section{Introduction}
Molecular dynamics has been proven to be a well suited method to
investigate the dynamic and static behavior of granular matter. The
concept of molecular dynamics was initially used to simulate the
dynamics of atoms and molecules, i.e.~of particles without inner
degrees of freedom. Alder and Wainwright who belong to the inventors
of this method investigated already 1957 numerically the low density
approximation of the entropy $S=-N~k_B~\langle \ln~f(p,t)\rangle$ ($f$
is the one--particle probability density in momentum space) in a hard
sphere system consisting of $N=100$
particles~\cite{AlderWainwright:1957ug}. There are applications of
molecular dynamics in many fields of interest and molecular dynamics
became one of the standard methods in computational physics. Recent
molecular dynamics simulations are of very high algorithmic
complexity, and systems consisting of up to $10^9$ particles
(with simple interaction forces) have been simulated (see
e.g.~\cite{BeazleyLomdahl:1993,BeazleyLomdahlJensenTamayo:1994,BeazleyEtAl:1995}). A
very interesting overview on the historical development is given in
the first part of Hoover's book~\cite{Hoover:1992}.

In molecular dynamics of granular particles, sometimes called
``granular dynamics'', each of the ``molecules'' is a macroscopic body
typically with a diameter of the order $D\approx 0.1\dots 1~mm$ which
has its own thermodynamic properties, i.e.~it has internal degrees of
freedom and hence it can dissipate mechanical energy when it is
subjected to outer forces. In some special cases as in simulations of
planetary rings~\cite{Salo:1992,Weidenschilling} the diameter can also
be of the order $D\approx 1~mm\dots 10~m$. In the most simple case one
can express the inelastic nature of the collisions of the grains by
introducing restitution coefficients which describe the relative
velocities in normal and tangential direction $\vec{v}_{ij}^{\,N}$ and
$\vec v_{ij}^{\,T}$ of two granular particles $i$ and $j$ after a
collision as functions of the relative velocities before the collision
$\vec V_{ij}^N$ and $\vec V_{ij}^T$
%\begin{mathletters}
\begin{eqnarray}
  \vec v_{ij}^{\,N} &=& -\epsilon_N\cdot \vec V_{ij}^N \hspace{0.5cm}
  \left( 0\le \epsilon_N \le 1\right) 
  \label{resitutionEQa}\\ 
   \vec v_{ij}^{\,T} &=&
    -\epsilon_T\cdot \vec V_{ij}^T \hspace{0.5cm} \left( -1\le
    \epsilon_T \le 1\right) ~.
  \label{resitutionEQb}
\end{eqnarray}
%\end{mathletters}
Here the relative velocity is given by
\begin{equation}
  \vec{v}_{ij} = \dot{\vec{r}}_i -\dot{\vec{r}}_j 
  \label{relvelocity1}
\end{equation}
or, including the particle rotation with angular velocities $\vec
\omega_i$ and $\vec \omega_j$ , by
\begin{equation}
  \vec v_{ij} = \left( \dot{\vec r}_i + \omega_i\times R_i \left(-\vec{n}\right)
\right) - \left( \dot{\vec r}_j + \omega_j\times R_j 
\vec{n} \right)
  \label{relvelocity2}
\end{equation}
with the unity vectors
%\begin{mathletters}
  \begin{eqnarray}
    \vec{n} &=& \frac{\vec r_i -\vec r_j }{\left| \vec r_i -\vec r_j
    \right|}\\ 
    \vec{t} &=& \left( \begin{tabular}{ll} 0 & -1 \\ 1 & 0
\end{tabular} \right) \cdot \frac{\vec r_i -\vec r_j }{\left| \vec r_i
  -\vec r_j \right|}~.
%    \label{unityvectors}
  \end{eqnarray}
%\end{mathletters}
Hence the relative velocities at the point of contact are
%\begin{mathletters}
  \begin{eqnarray}
    \vec v_{ij}^{\,N} &=& \vec n \left(\vec v_{ij} \cdot \vec n \right)\\
    \vec v_{ij}^{\,T} &=& \vec t \left(\vec v_{ij} \cdot \vec t \right) ~.
%    \label{relvelocities}
  \end{eqnarray}
%\end{mathletters}
In general the normal and tangential restitution coefficients $
\epsilon_N $ and $ \epsilon_T $ are themselves functions of the
relative impact velocities $\epsilon_N=\epsilon_N\left(
V_{ij}^N,~V_{ij}^T\right) $ and $\epsilon_T=\epsilon_T\left(
V_{ij}^N,~V_{ij}^T\right) $. For the case that the particles are rough
homogeneous spheres the coefficients $\epsilon_N$ and $\epsilon_T$ can
be calculated
analytically~\cite{BrilliantovSpahnHertzschPoeschel:1994} solving the
viscoelastic equations of the particles. Using this model one neglects
the complicated interaction between particles during the collision.
The approximation is justified if collisions are relatively rare
events, i.e. if the mean free path between collisions is long and one
can restrict the calculation to two--particle--interactions only. If a
system is in this regime one calls it a ``granular gas''. Event driven
algorithms have been applied in many papers, recently e.g.
in~\cite{GoldhirschZanetti:1993,McNamaraYoung:1992} for an
investigation of inelastic clustering of granular particles (using
constant restitution coefficients) and by Luding et al.~to simulate
vertically shaken
material~\cite{LudingClementBlumenRajchenbachDuran:1994DISS,LudingClementBlumenRajchenbachDuran:1994STUDIES}.
For sophisticated implementations of such algorithms
see~\cite{Rapaport:1980,Lubachevsky:1991}.

In many cases of interest, however, the particles collide very often
or they touch each other permanently, i.e. the system reveals static
properties. Examples for such systems are flows through pipes and
hoppers, ball mills, shaken materials, sand heaps and many others. In
the model proposed by Cundall and Strack~\cite{CundallStrack:1979} and
Haff and Werner~\cite{HaffWerner:1986} the particles feel restoring
forces in normal and tangential direction, and they lose kinetic
energy during the collisions. If the particles $i$ and $j$ at the
positions $\vec{r}_i$ and $\vec{r}_j$ with radiuses $R_i$ and $R_j$
touch each other they feel the force
%\begin{mathletters}
\begin{equation}
  \vec{F}_{ij}= F_{ij}^N\, \vec{n} + F_{ij}^T\, \vec{t}
\end{equation}
with 
\begin{eqnarray}
  F_{ij}^N &=& Y \cdot \left(R_i+R_j- \left| \vec{r}_i-\vec{r}_j\right| 
  \right)^{\frac{3}{2}} - m_{ij}^{eff} \cdot \gamma_N \cdot 
  (\dot{\vec{r}}_i - \dot{\vec{r}}_j) \cdot\vec{n}
\label{fnormal}\\ 
F_{ij}^T &=& \mbox{sign} (v_{ij}^{rel}) \cdot \min\left( m_{ij}^{eff} \gamma_T \left|
v_{ij}^{rel}\right| , \mu \left| F_{ij}^{N}\right| \right) 
\label{ftang} \\
v_{ij}^{rel} &=& (\dot{\vec{r}}_i - \dot{\vec{r}}_j)  \cdot
\vec{t}+ R_i \cdot \Omega_i + R_j \cdot \Omega_j 
\label{surfvelocity}\\ 
m_{ij}^{eff} &=& \frac{m_i \cdot m_j}{m_i + m_j} ~.
\label{meff}
\end{eqnarray}
%\end{mathletters}
$Y$ is the Young modulus, $\gamma_N$ and $\gamma_T$ are the damping
coefficients in normal and tangential direction and $\mu$ is the
Coulomb friction coefficient. Eq.~(\ref{surfvelocity}) describes the
relative velocity of the surfaces of the particles at the point of
contact and eq.~(\ref{meff}) gives the effective mass.
Eq.~(\ref{ftang}) takes the Coulomb friction law into account, saying
that two particles slide on top of each other if the shear force
overcomes $\mu$ times the normal force. Eq.~(\ref{fnormal}) comes
from the Hertz law \cite{Hertz:1882ug} for the force between two rigid
spheres which are in contact with each other.

This model has been successfully applied to simulate the behavior of
dry granular material in many works describing various physical
phenomena. It developed to the standard method for calculations of
granular dynamics. Some of these problems are size segregation and
convection in vibrated material in two and three dimensions
(e.g.~\cite{GallasHerrmannSokolowski:1992PRL,Taguchi:1992PRL,GallasHerrmannPoeschelSokolowski:1994,PoeschelHerrmann:1994}),
the flow in hoppers
(e.g.~\cite{RistowHerrmann:1994,FormKohringMelinPuhlTillemans:1994,CampbellPotapov:1993})
and pipes (e.g.~\cite{Poeschel:1994,Lee:1994}), the flow in rotating
cylinders
(e.g.~\cite{Ristow:1994,RistowCantelaubeBideau:1994ug,PoeschelBuchholtz:1993CSF}),
the motion of granular material on an inclined
surface~\cite{Walton:1991,Poeschel:1993JDP}, sound propagation in
granular material~\cite{Melin:1994}, the onset of
turbulence~\cite{Taguchi:1992JDP,Taguchi:1994} and many others.  Many
experimental results of many authors could be reproduced numerically
using this type of molecular dynamics simulations. There have been
developed very efficient algorithms for the case of short range
interaction with large force gradients as it is typical for granular
materials,
e.g.~\cite{BuchholtzPoeschel:1993JMPC,FormItoKohring:1993,EveraersKremer:1993}.
The model by Cundall and Strack yields reliable results in more
dynamic systems, i.e. when the static friction of the particles does
not play a major role. When the static properties of the material
govern the behavior, however, the model might fail.
In~\cite{BuchholtzPoeschel:1994PA} is shown that one cannot build up a
stable sand heap with a finite inclination of such particles but the
heap dissolves under its own mass and the angle becomes smaller with
rising number of particles.  Lee~\cite{Lee:1993} simulated static
friction effects using spheres which are connected by springs. When
two particles touch each other for a certain time a spring ``grows''
between these particles and keeps the particles together when forces
are applied which would separate them. When the separating forces
become too large the springs break and the particles move freely.
Using this model Lee reproduced the finite angle of repose in a sand
heap, however, the growing and breaking springs have no direct
equivalent in nature, and hence the model seems to be quite
artificial.

Another simulation method which corresponds to the previous one for
the case of very large damping has been introduced by Visscher and
Bolsterli~\cite{VisscherBolsterli:1972ug}. In each iteration step all
of the grains are moved according to the motion of a vessel containing
the granular material. Then the particles are released one by one,
beginning with the lowest one (with respect to the direction of
gravity). Each particle moves until it reaches the next local minimum
of the potential energy when it hits either the wall or other
previously released grains. When a particle reached its local minimum
it stays in that position, i.e.~a falling
particle cannot cause the motion of another previously released
particle. This behavior means that inertia of the moving particles is
neglected as soon as they hit another particle or the wall, and hence
it corresponds to the case of very large damping. The advantage of
this algorithm is its numerical simplicity, i.e.~it is possible to
investigate much larger systems than using molecular dynamics.
Recently it has been applied in simulations of several problems as
size
segregation~\cite{JullienMeakinPawlovitch:1992}
and the motion of particles in a rotating
cylinder~\cite{BaumannJobsWolf:1993,BaumannJanosiWolf:1994}. Obviously
the simplification of infinite damping is not valid in each case, and
hence one has to carefully investigate whether it is justified to apply
this algorithm (see
e.g.~\cite{BarkerMehta:1995}).

There are some more simulation methods for the dynamics of granular
material which we want to mention only: Peng and Herrmann applied a
Lattice Gas Automaton to investigate density waves in a
pipe~\cite{PengHerrmann:1994}. Caram and Hong investigated granular 
flows using a Random Walk technique~\cite{CaramHong:1991}.

There are some models to simulate complex shaped grains which are
composed of spheres. In the model by Gallas and Soko{\l}owski
\cite{GallasSokolowski:1993} the grains consist of two spheres
connected with each other by a stiff bar. Walton and Braun applied
more complicated particles consisting of four or eight spheres rigidly
connected with each other~\cite{WaltonBraun:1993}. Using this model
they examined the transition from stationary to sliding flow and the
transition from sliding to raining flow in a rotating drum. P\"oschel
and Buchholtz~\cite{PoeschelBuchholtz:1993,BuchholtzPoeschel:1994PA}
describe grains built up of five spheres where one of them is located
in the center of the grain and four identical spheres are at the
corners of a square. Each pair of neighboring spheres is connected by
a damped spring. The latter model was applied to the rotating cylinder
\cite{PoeschelBuchholtz:1993CSF} and it was shown that the simulation
results agree much better with the experiment than equivalent
calculations with spheres. Especially it was shown that one can
reproduce stick--slip motion and avalanches which was not possible in
simulations using spheres. The inclination of the surface of the
material in the rotating cylinder and the dependency of the
inclination on the angular velocity, however, did not agree well with
the experiment.

Mustoe and DePorter~\cite{MustoeDePorter:1993} proposed a particle
model where the boundary geometry of the particles is defined (in
local coordinates) by
\begin{equation}
f_i\left( x_i,y_i\right) =\left[\frac{\left| x_i \right|}{a_i}\right]^{n_i} + 
             \left[\frac{\left| y_i \right|}{b_i}\right]^{n_i} - 1 = 0.
 \label{hogue.eq}
\end{equation}
Varying $n_i$ between 2 and $\infty$ the shape of the $i$th particles
varies continuously from elliptical to rectangular. To detect contacts
of pairs of particles described by eq.~(\ref{hogue.eq}) one has to
solve numerically equations of high order for each pair of possibly
touching particles in each iteration step. This requires an iteration
technique which converges slowly for higher exponents $n_i$ and which
makes hence the algorithm numerically complicated. Hogue and
Newland~\cite{HogueNewland:1993} investigated the flow of granular
material on an inclined chute and through a hopper using a model where
the boundaries of the convex particles are given by polygons with up
to 24 vertices each. To detect whether two particles touch each other
one has to calculate the intersections between each pair of vertices.
During collisions energy is dissipated according to Stronge's energy
dissipation hypothesis~\cite{Stronge:1990} in normal direction, whilst
Coulomb's friction law models the energy losses also in tangential
direction.

Using grains which consist of interconnected spheres or of particles
described by eq.~(\ref{hogue.eq}) it is not possible to simulate
particles with sharply formed corners. For some effects it seems to be
essential to simulate such particles to reproduce the experimental
observed effects. This point is discussed in detail
in~\cite{BuchholtzPoeschel:1994PA}. Tillemans and
Herrmann~\cite{TillemansHerrmann:1994,Tillemans:1994} proposed a two
dimensional particle model where the particles are convex polygons.
When two particles $i$ and $j$ touch each other, i.e. when there is an
overlap, there acts the force
%\begin{mathletters}
\begin{eqnarray}
  \vec{F}_{ij}&=&F_{ij}^N~\vec{e}^{\,N} + F_{ij}^T~\vec{e}^{\,T}\\
  F_{ij}^N &=& \frac{Y~A}{L_c} -
  m_{ij}^{eff}~\gamma_N~\left( \vec{v}^{\,rel}\cdot \vec{e}^{\,N} \right) \\ 
  F_{ij}^T &=& - \min\left( m_{ij}^{eff} \gamma_{T}
  \left|\vec{v}_{rel} \cdot \vec{e}^{\,T} \right|, \mu \left|
    F_{ij}^{N} \right|\right)~.
%  \label{forcetillemans}
\end{eqnarray}
%\end{mathletters}
$A$ is the compression area of the particles
(overlap area), Y is the Young module of the material.
The effective mass $m_{ij}^{eff}$ is given by eq.~(\ref{meff}), the
relative particle velocity is $\vec{v}^{\,rel}= \dot{\vec{r}}_i
-\dot{\vec{r}}_j $, where $\vec{r}_i$ points to the center of mass
of the particle $i$, and $L_c$ is the characteristic size of the
particle. The unit vectors are in the direction of the line that
connects the intersection points of the overlapping particles $i$ and
$j$ ($\vec{e}^{\,T}$) and perpendicular to this line
($\vec{e}^{\,N}$). Only convex particles are allowed. This model was
used e.g.~in simulations of shear cells, earth quakes and flow through
hoppers~\cite{Tillemans:1994}. The results differ significantly from
similar calculations using spheres. For the case of the hopper
simulation they found clogging and arching which could not be found in
simulations with spheres. A similar model was used earlier by
Handley~\cite{Handley:1993} who investigated the fracture behavior of
brittle granular material. Potapov et
al.~\cite{PotapovCampbellHopkins:1994aug,PotapovCampbellHopkins:1994bug}
investigated a similar model for solid fracture. Initially they
subdivide a macroscopic two dimensional body into many small
equilateral triangles (elements). The forces in the body are resolved
as forces at inter--element contacts of two different types which they
call ``glued'' and ``collisional''. Glued contacts are the joints
between elements interior to a solid body which can support stresses.
When the tensile stresses exceed a certain value, cracks form and the
glued bond breaks. Collisional contacts are contacts between the
surface elements of the body (which eventually collides with other
bodies or walls), and contacts between triangles in the bulk of the
body where glued contacts have been broken. In the three latter
models~\cite{TillemansHerrmann:1994,Tillemans:1994,Handley:1993,PotapovCampbellHopkins:1994aug,PotapovCampbellHopkins:1994bug}
the convex grains have been Voronoi polyhedra in two
dimensions~\cite{Finney:1979}.

In this paper we present a new particle model for molecular dynamics
of granular material in two dimensions where the particles are
simulated using a similar model which has some advantages. In the
following section we describe the model, the forces acting between the
grains are derived in sections~\ref{trianglesSEC} and
\ref{stressesinbeamsSEC}. Some aspects of the implementation and the
performance of the algorithm are discussed in
section~\ref{implementationSEC} and some sample results based on this
particle interaction model will be briefly discussed in
section~\ref{resultsSEC}.

\section{The model}
\begin{figure}[ht]
  \ifcase \value{nofigs}
  \begin{center}
    \begin{minipage}{4cm}
      \psfig{figure=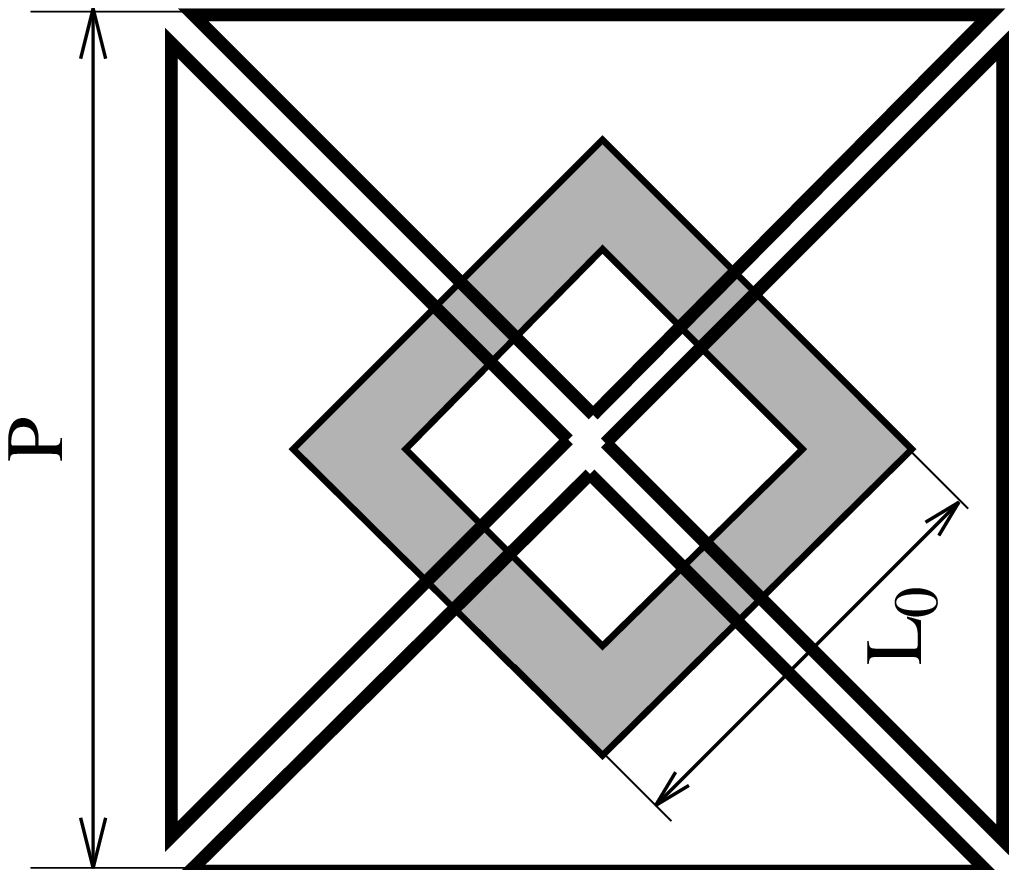,width=4cm,angle=270}
    \end{minipage}
    \hspace{2cm}
    \begin{minipage}{2cm}
      \psfig{figure=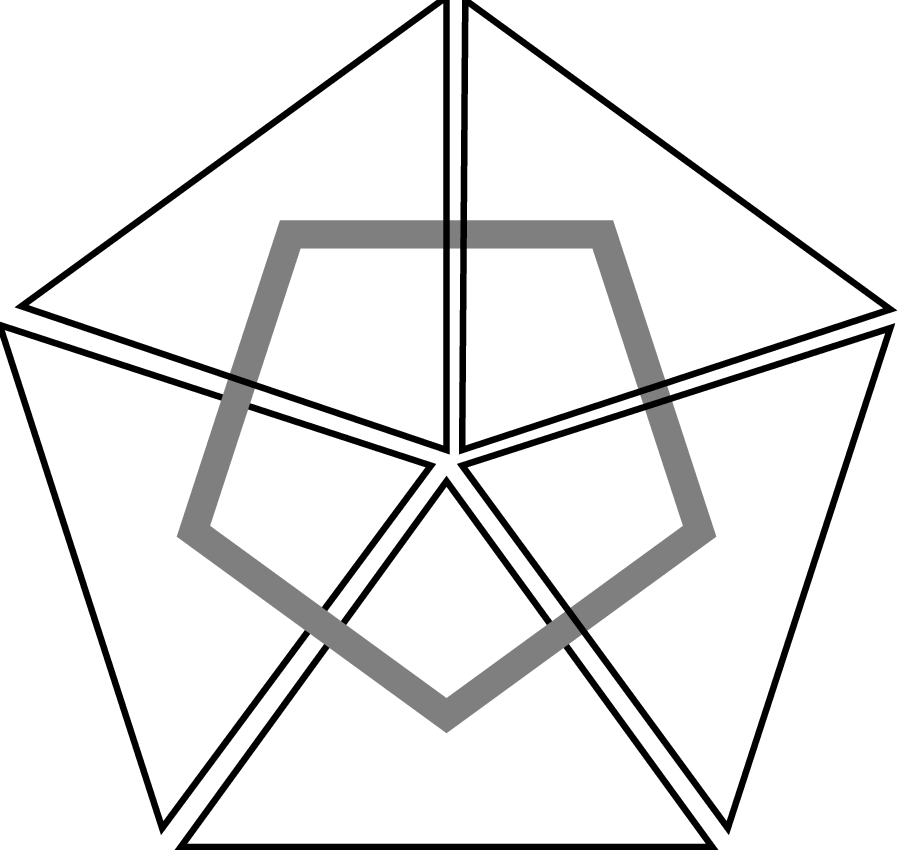,width=2cm,angle=270} 
      \vspace{0.5cm}
      
      \psfig{figure=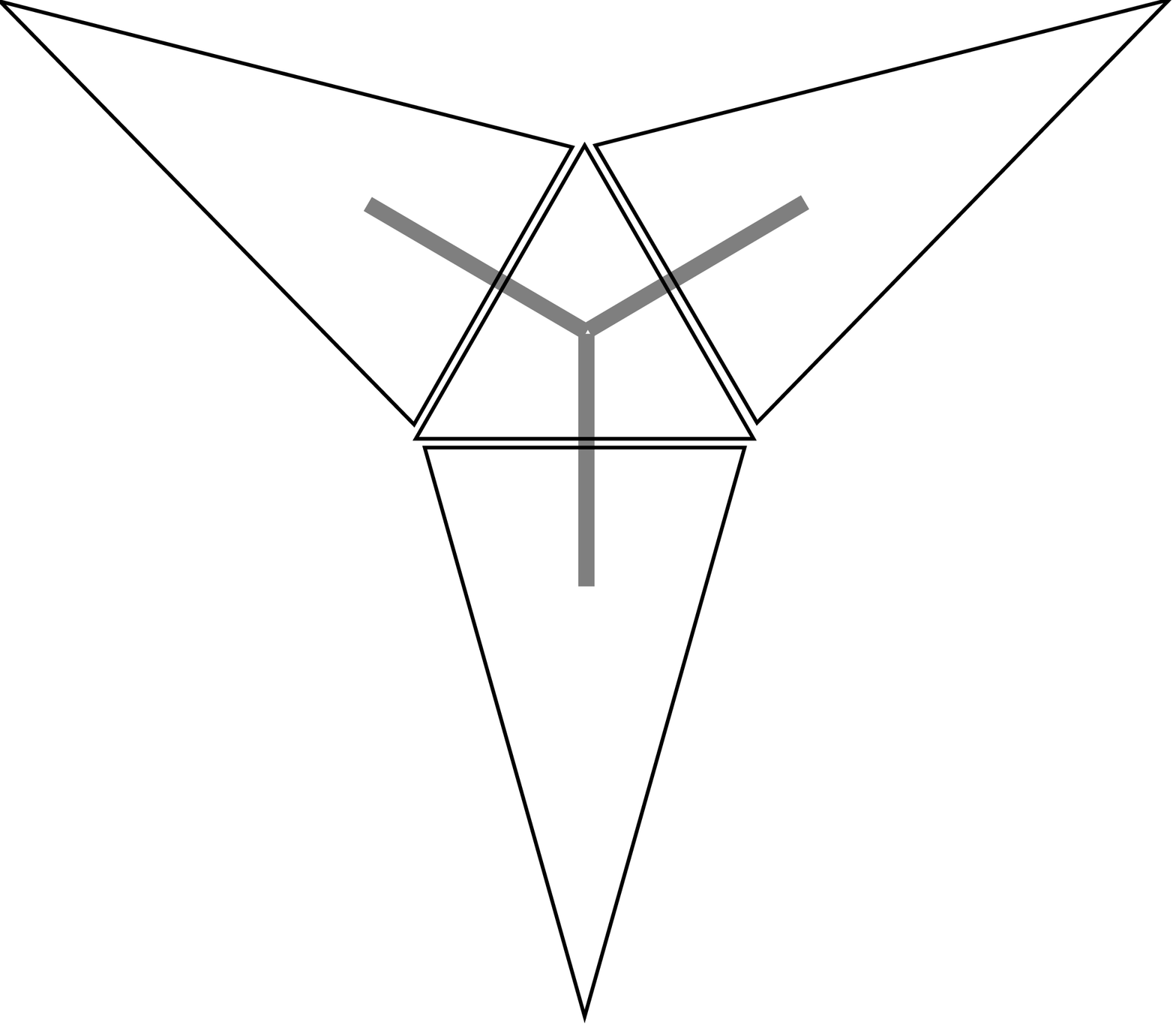,width=2cm,angle=270}
    \end{minipage}
  \end{center}
  \vspace{0.5cm}
  
  \fi
  \caption{Examples of grains composed of different numbers of triangles. 
    The model is not restricted to convex grains.}
  \label{grainsFIG}
\end{figure}

In our model the grains are composed of an arbitrary number of
ideal elastic triangles which are connected by beams (see
fig.~\ref{grainsFIG}). 

A beam in our sense is an deformable damped bar
which is subjected to forces in the direction of its axis and
transverse to its axis, i.e.~to normal and shear forces, and to
moments acting on its ends. Triangles which belong to the same grain
do not interact with each other. When two triangles of different
grains collide, i.e.~if there is an overlap between both, they feel a
restoring force. Hence the force acting upon the triangle $i$
belonging to the grain $j$ ($j\in
\{1,2,\dots,N\}$) is
\begin{equation}
\vec{F}_i^{\,j} = \sum_{l=1,l\neq j}^N 
      \sum_{k=1}^{n_t(l)} \vec{\Gamma}_{ik}^{jl} 
   + \sum_{k=1}^{n_b^j(i)} \vec\Lambda_{ik}^{\,j} 
   + \vec\Phi(\vec{r}_{i}^{\,j}, \dot{\vec{r}}_i^{\,j})~.
  \label{totalforceEQ}
\end{equation}
The first sum in the first term in eq.~(\ref{totalforceEQ}) runs over
all grains $l$ except the $j$th, the second sum runs over all
triangles $k\in\{1,2,\dots,n_t(l)\}$, the $l$th grain consists of.
$\vec{\Gamma}_{ik}^{jl}$ is the force which acts between the triangles
$i$ and $k$ which belong to different grains $j$ and $l$ (see section~\ref{trianglesSEC}). The sum in
the second term runs over all beams connecting the triangle $i$ of the
grain $j$ with other triangles of the same grain.
$\vec\Lambda_{ik}^{\,j}$ is the force induced by the beam $k$ acting
on the $i$th triangle of the grain $j$. They originate from the
distortion of the beam that connects the triangle $i$ with another one
of the same grain. The acting forces and momenta will be discussed
below in section~\ref{stressesinbeamsSEC}. A similar model for the
beam forces has been introduced earlier by Herrmann et al. for the
fracture of disordered lattices~\cite{HerrmannHansenRoux:1989}.  The
last term $\vec\Phi(\vec{r}_{i}^{\,j}, \dot{\vec{r}}_i^{\,\,j})$
describes the action of an external force, as e.g.~gravity, on the
triangles.

During its deformation a beam dissipates energy similar to a linearly
damped spring, i.e.~proportional to its deformation rate~(see
section~\ref{stressesinbeamsSEC}). To avoid time consuming
calculations due to Steiner's law each beam is fixed at both ends in
the center of mass points of the triangles it connects. When the beam
is not deformed by eventually applied forces (the beam is at rest) it
is perpendicular to the neighboring edges of the triangles which it
connects.  Fig.~\ref{grainsFIG} shows some examples of grains. The
only restriction concerning the number, the shape and the position of
the triangles is caused by the condition that the beams are fixed in
the center of mass.

The next section~\ref{trianglesSEC} gives a detailed description of the
force that acts when triangles collide, in
section~\ref{stressesinbeamsSEC} the forces due to deformed beams are
derived.

\section{The interaction of the triangles}
\label{trianglesSEC}
\begin{figure}[ht]
  \ifcase \value{nofigs}
\centerline{\psfig{figure=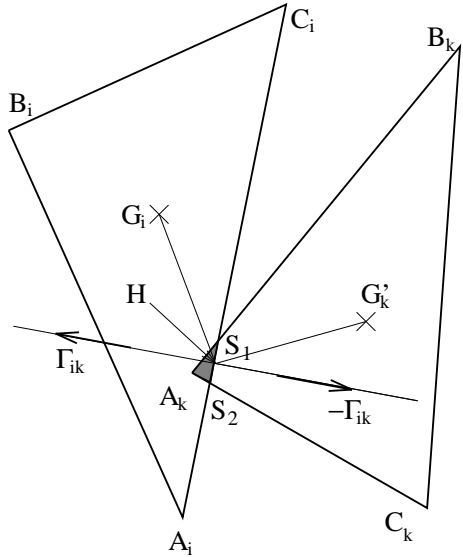,width=6cm,angle=270}}
\vspace{0.5cm}
\fi
\caption{The ``standard'' type of a collision 
  between two triangles $A_iB_iC_i$ and $A_kB_kC_k$. The forces
  $\pm\vec{\Gamma}_{ik}$ are directed perpendicular to the
  intersection line $\overline{S_1S_2}$.  Moments act with respect to
  the center of mass points $G_i$ and $G_j$.  The absolute value of
  the interaction force $\left|\vec\Gamma_{ik}\right|$ is proportional
  to the shadowed intersection area $\Delta \left(S_1S_2A_k\right)$.}
\label{standard.fig}
\end{figure}

In this section we describe the calculation of the term
$\vec{\Gamma}_{ik}^{jl}$ in eq.~(\ref{totalforceEQ}) originating from
the compression of two triangles $i$ and $k$ which belong to the
grains $j$ and $l$, respectively. When the triangles are compressed
there exists a virtual ``overlap area'' which leads to an elastic restoring
force $\vec{\Gamma}_{ik}$. For simplicity here and in the following we
drop the upper indexes of the variables. Areas of triangles $XYZ$ and
quadrangles $WXYZ$ we denote by $\Delta(XYZ)$ and $\Box(WXYZ)$.
Fig.~\ref{standard.fig} shows the most frequently found type of a
collision between the triangles $A_iB_iC_i$ and $A_kB_kC_k$, i.e.~a
corner of one triangle deforms a side of another one. 

The absolute
value of the restoring force $\vec{\Gamma}_{ik}$ is given by the
shadowed area $\Delta\left(S_1S_2A_k\right)$ of the triangle
$S_1S_2A_k$ times the Young module $Y$. Its direction is
perpendicular to the intersection line $\overline{S_1S_2}$ (Poisson
hypothesis, see e.g.~\cite{FormKohringMelinPuhlTillemans:1994}). Hence
resulting momenta acting upon the triangles $A_iB_iC_i$ and
$A_kB_kC_k$ read
%\begin{mathletters}
\begin{eqnarray}
\vec{M}\left(A_iB_iC_i\right) &=& \vec{HG_i} \times \vec{\Gamma}_{ik} \label{momentaA}
\\
\vec{M}\left(A_kB_kC_k\right) &=& \vec{HG_k} \times \vec{\Gamma}_{ki} = 
  -\vec{HG_k}\times \vec{\Gamma}_{ik}~.
\label{momentaB}
\end{eqnarray}
%\end{mathletters}
The vector $\vec{HG_i}$ points from the middle point of the line
$\overline{S_1S_2}$ to the center of mass of the triangle $A_iB_iC_i$.

Although the case shown in fig.~\ref{standard.fig} is the mostly
occurring type of collisions there are several other types which
treatment has to be discussed below.

\begin{figure}[ht]
  \ifcase \value{nofigs}
\centerline{\psfig{figure=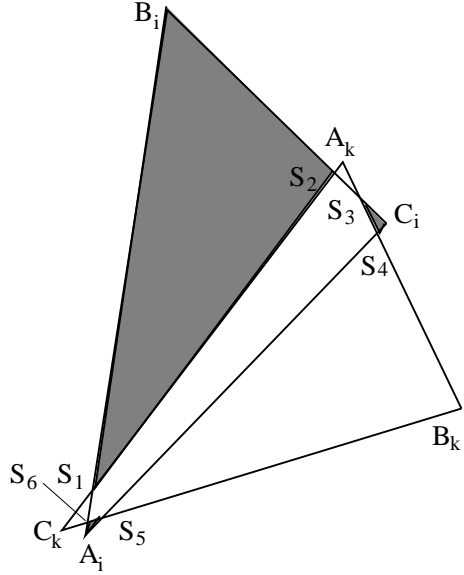,width=6cm,angle=270}}
\fi
\caption{The fifth type of collisions: There are six intersection
points $S_1$,\dots,$S_6$.\hspace{2cm}}
\label{schnit5}
\end{figure}

Classifying the possible interactions of the triangles we define five
different types of collisions which are slightly different from those
in~\cite{PotapovCampbellHopkins:1994aug}. The detection of overlapping
polygons is a very important problem in computer graphics too, where
one frequently has to decide whether two objects cover each other, and
which of them is in the foreground or in the background, respectively.
Therefore we found it helpful and inspiring first to have a look into
advanced methods in computer graphics,
e.g.~\cite{BentleyOttmann:1979ug,PreparataShamos:1985ug}. In all
figures~\ref{standard.fig}--\ref{schnit5} the overlapping area is
drawn extremely exaggerated. Typically the overlapping area of two
interacting triangles does not exceed a few tenth of a percent of the area
of the triangles.

\begin{enumerate}
\begin{figure}[ht]
  \ifcase \value{nofigs}
  \centerline{\psfig{figure=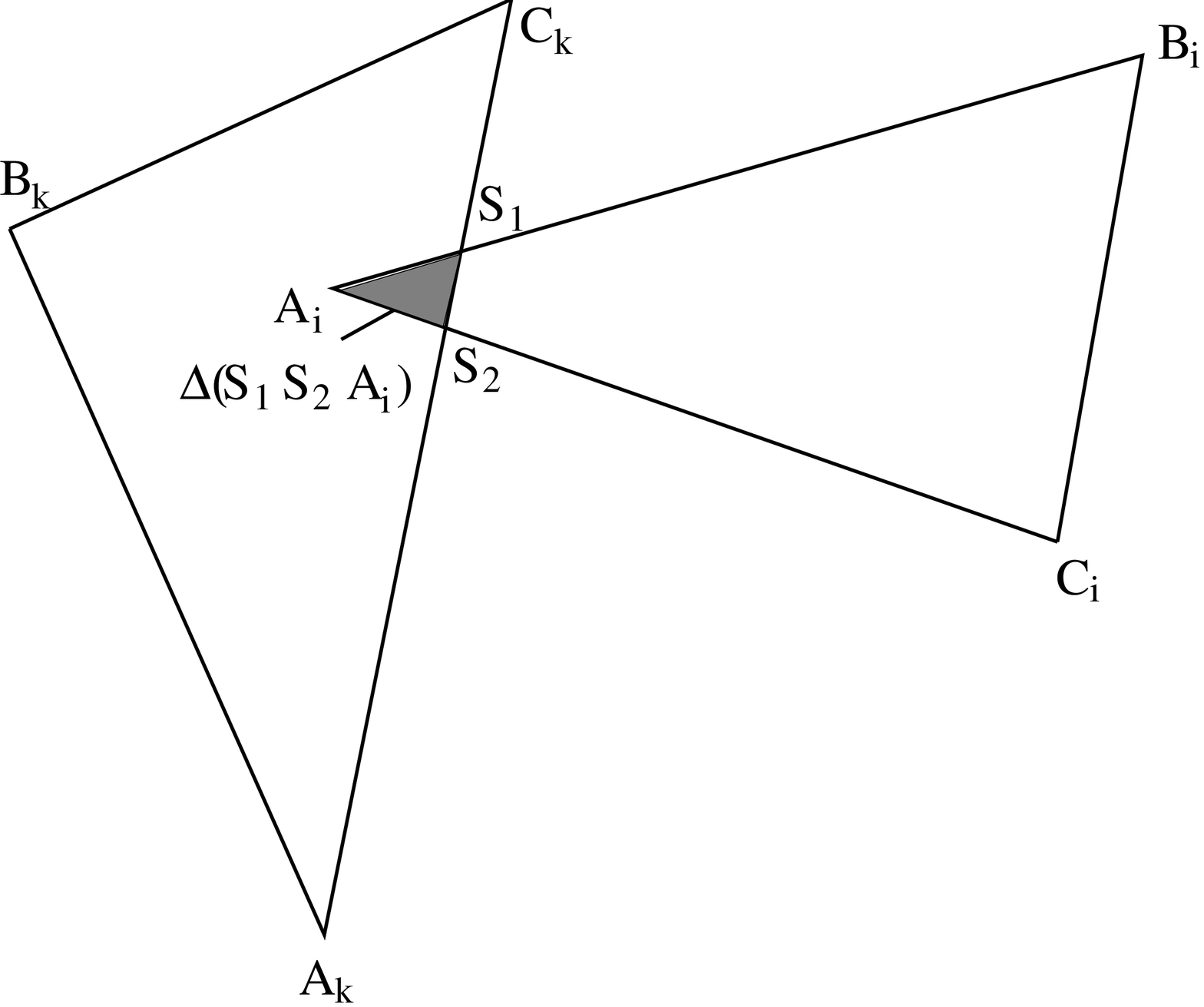,width=6cm,angle=270}\psfig{figure=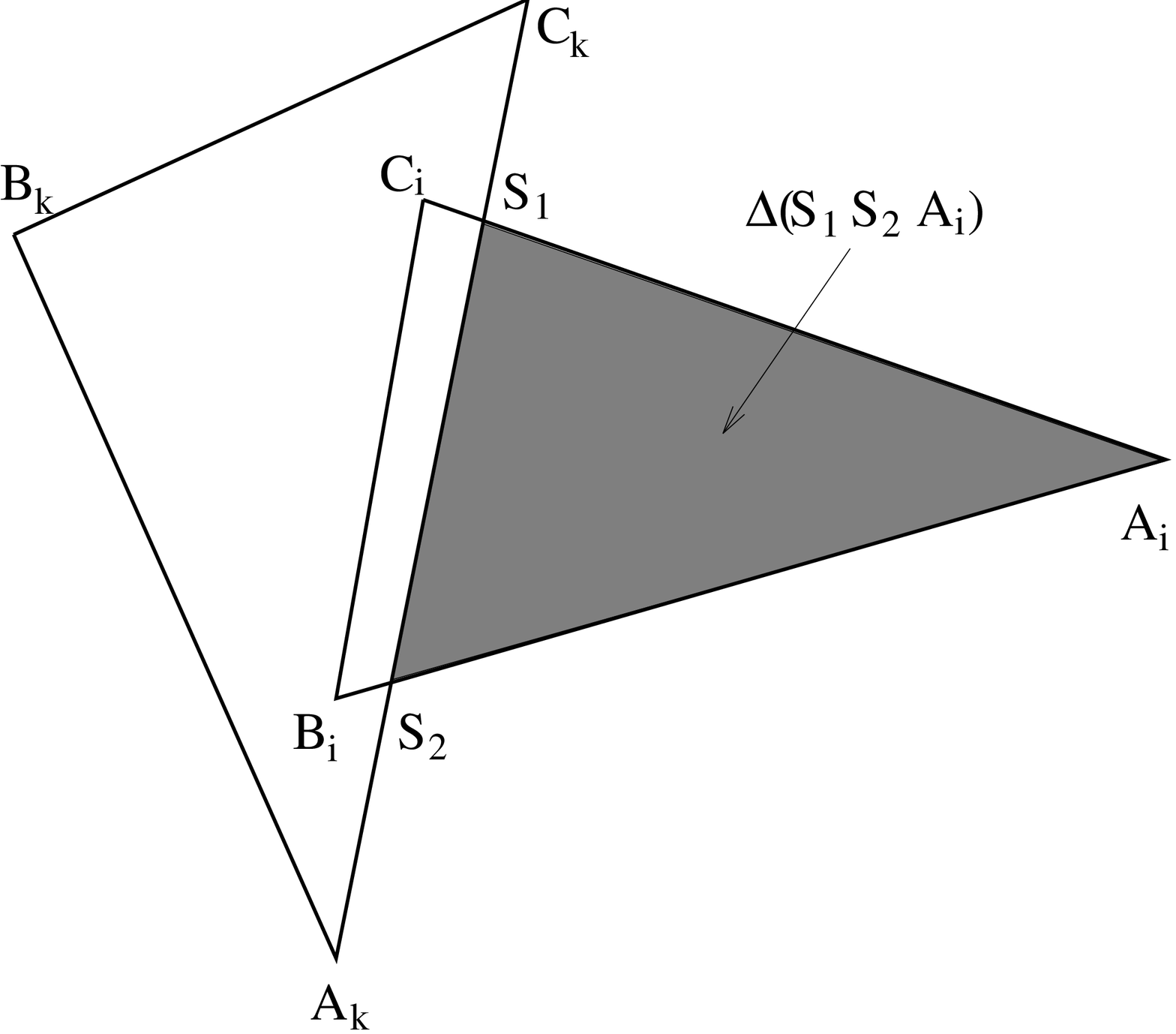,width=6cm,angle=270}}
  \vspace{0.5cm}
\fi
\caption{The first type of collisions: There are two intersection 
points $S_1$ and $S_2$. Both lie at the same edge of one of the 
triangles.}
\label{schnit1}
\end{figure}

\item{For the first type (fig.~\ref{schnit1}) there are two
    intersection points $S_1$ and $S_2$ lying at the same edge of one
    of the interacting triangles. One of the situations drawn in
    fig.~\ref{schnit1} corresponds to the standard collision
    (fig.~\ref{standard.fig}). We calculate the area $\Delta \left( S_1
      S_2 A_i\right)$ (shadowed area) given by the intersection points
    $S_1$ and $S_2$ and the included point $A_i$. Because the
    overlapping area is definitely smaller than half the area of the
    triangles the force is proportional to the minimum between $\Delta
    \left( S_1 S_2 A_i\right)$ and $\Delta \left( A_i B_i C_i\right) -
    \Delta \left( S_1 S_2 A_i\right)$
\begin{equation}
  \left|\vec{\Gamma}_{ik}\right| = \Gamma_{ik} = 
  Y \cdot \min \left\{\Delta \left( S_1 S_2 A_i\right) , \Delta \left(
  A_i B_i C_i\right) - \Delta \left( S_1 S_2 A_i\right) \right\}~.
\end{equation}
The force acts perpendicular to the line between the intersection
points $\overline{S_1 S_2}$.}

\item The second type (fig.~\ref{schnit2}) is characterised by two
intersection points lying on different edges for both triangles. We
calculate the areas $\Delta \left( S_1 S_2 A_i\right) $ (dark gray)
and $\Delta \left( S_1 S_2 A_k\right) $ (light gray). The force is
proportional to the overlap
\begin{eqnarray}
\left|\vec{\Gamma}_{ik}\right| = \Gamma_{ik} = Y \cdot \left[
\min \left\{\Delta \left( S_1 S_2 A_i\right) , \Delta 
\left( A_i B_i C_i\right) - \Delta \left( S_1 S_2 A_i\right) 
\right\} + \right.\nonumber \\
\left. \min \{\Delta \left( S_1 S_2 A_k\right) , \Delta 
\left( A_k B_k C_k\right) - \Delta \left( S_1 S_2 A_k\right) \}
\right]~.
\end{eqnarray}
It acts perpendicular to the line between the intersection points
as in the previous case.
\begin{figure}[ht]
  \ifcase \value{nofigs}
\centerline{\psfig{figure=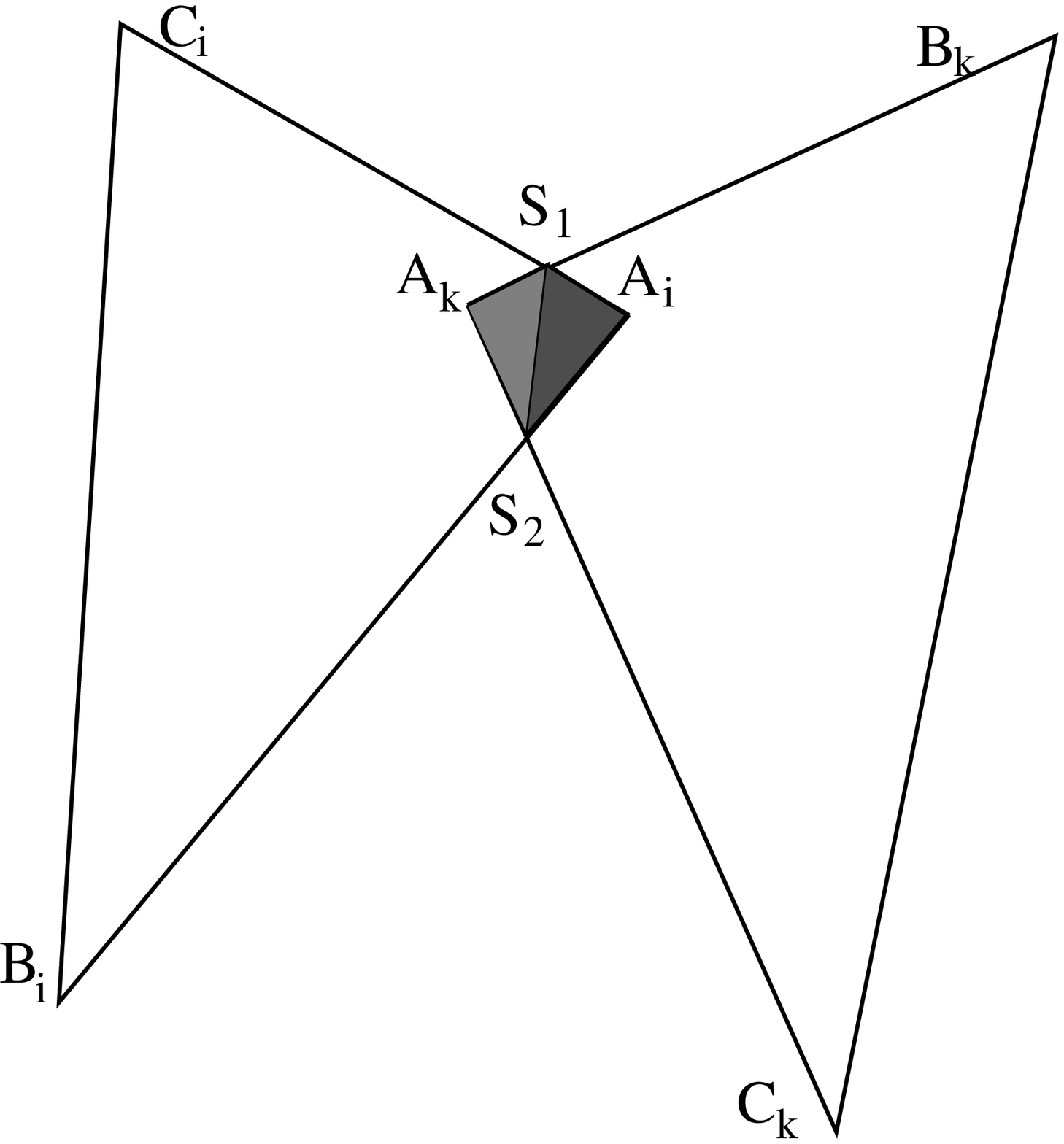,width=6cm,angle=270}\hspace{0.5cm}\psfig{figure=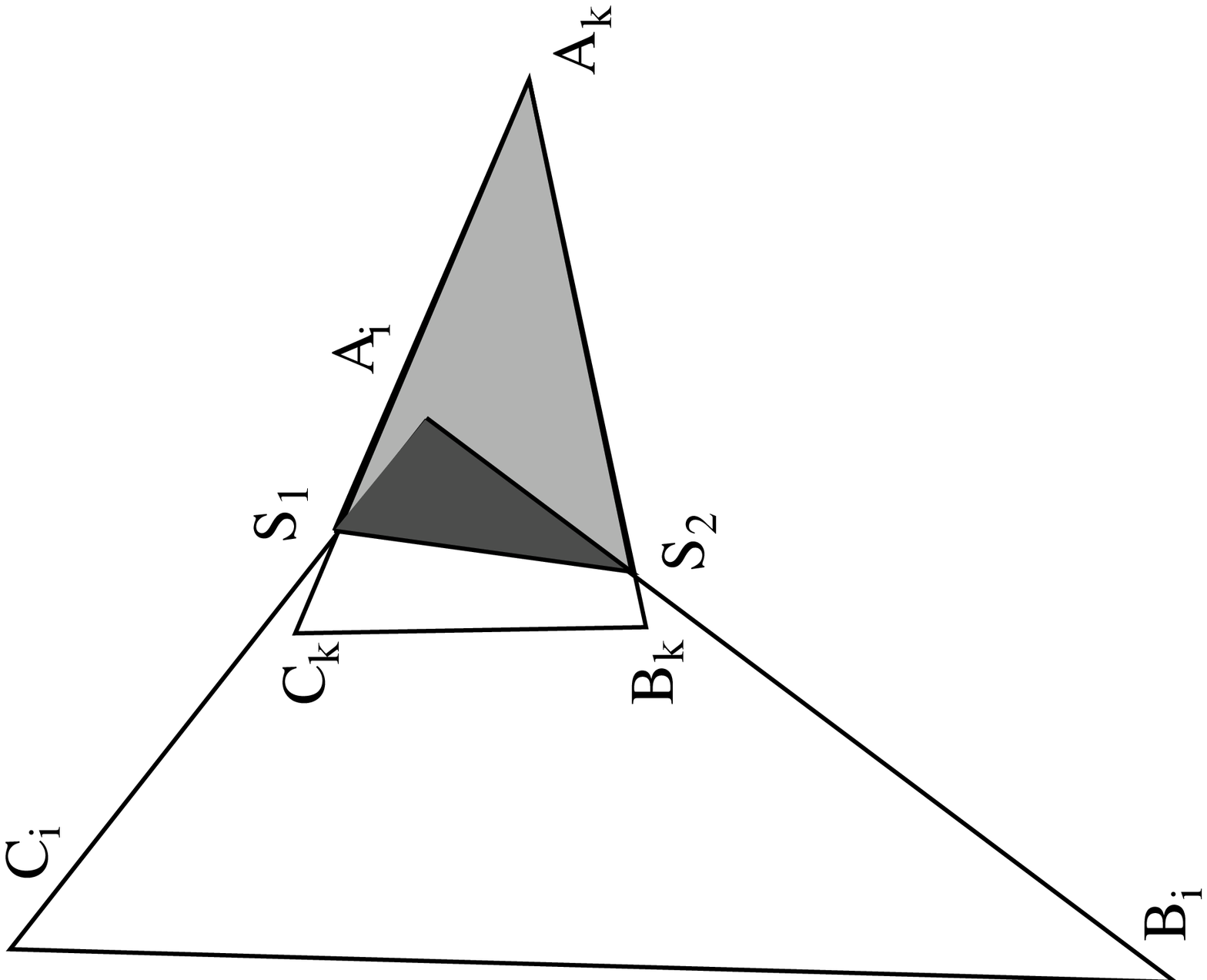,width=6cm,angle=270}}
\vspace{0.5cm}
\fi
\caption{The second type of collisions: There are two intersection 
points $S_1$ and $S_2$ lying on different edges for both triangles.}
\label{schnit2}
\end{figure}

\item {The third type (fig.~\ref{schnit3}) is characterised by four
    intersection points, each pair of two points lies on one edge of
    both triangles. Therefore the third edge of both interacting
    triangles does not intersect the edges of the other triangle.
    There are two pairs of forces proportional to the quadrangular
    area $\Box \left( S_1 S_2 S_3 S_4\right) $ of the overlap
\begin{equation}
\left|\vec{\Gamma}_{ik}\right| = \Gamma_{ik}  = 
\frac{1}{2} \cdot Y \cdot \Box \left( S_1 S_2 S_3 S_4\right)~, 
\end{equation}
one acting perpendicular to the line $\overline{S_1 S_2}$ and the
other perpendicular to the line $\overline{S_1 S_4}$. The pre--factor
$0.5$ is due to two pairs of acting forces instead of one for the first two
collision types.}
\begin{figure}[ht]
  \ifcase \value{nofigs}
\centerline{\psfig{figure=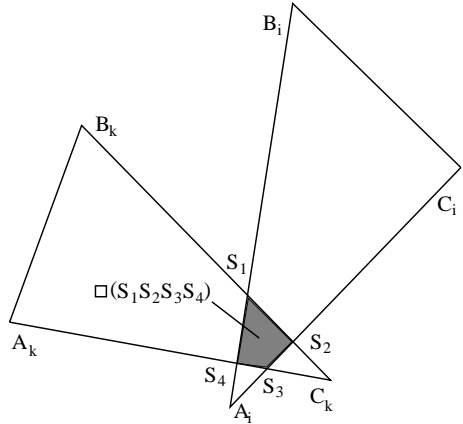,width=6cm,angle=270}}
\vspace{0.5cm}
\fi
\caption{The third type of collisions: There are four intersection 
points $S_1$, $S_2$, $S_3$ and $S_4$. Each pair of points lies on 
one edge of either the first or the second triangle.}
\label{schnit3}
\end{figure}

\item{The fourth type (fig.~\ref{schnit4}) is characterised by four
    intersection points too, but here one of the triangles has
    intersection points at all its edges. Because interactions of this
    type occur extremely seldomly we did not implement an exact
    calculation for this case, but we calculate two interactions of
    type $1$ with the intersection points $S_1$--$S_2$ or $S_3$--$S_4$
    respectively, instead of solving the correct problem. In real
    simulation the fraction of this type occurs approximately
    $10^{-5}$ of the number of collisions.  Nevertheless one has to
    deal with these extremely seldom events since otherwise they might
    cause problems from which the system cannot recover.  If one does
    not care about these events usually the system gains in a few time
    steps after the event occurs a huge amount of kinetic energy,
    i.e.~the system explodes.}
\begin{figure}[ht]
  \ifcase \value{nofigs}
\centerline{\psfig{figure=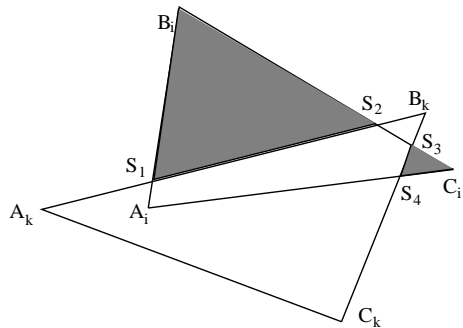,width=6cm,angle=270}}
\vspace{0.5cm}
\fi
\caption{The fourth type of collisions: There are four intersection 
points $S_1$, $S_2$, $S_3$ and $S_4$. One of the triangles intersects 
with all its three edges.}
\label{schnit4}
\end{figure}

\item{The fifth and last type of interaction (fig.~\ref{schnit5}) is
characterised by six intersection points $S_1$ -- $S_6$. Equivalent to
the forth type we do not calculate the exact interaction but
substitute this calculation by solving three interactions of the first
type with pairs of interaction points $S_1$--$S_2$, $S_3$--$S_4$ and
$S_5$--$S_6$}.
\end{enumerate}
According to the forces $\vec{\Gamma}_{ik}$ we derive forces in
parallel to the axes of the Cartesian coordinate system and moments
$\vec{M}$ acting with respect to the center of mass points $G_i$ and
$G_k$ as described in equations~(\ref{momentaA}, \ref{momentaB}).

\section{Stresses in beams}
\label{stressesinbeamsSEC}
\begin{figure}[ht]
  \ifcase \value{nofigs}
\centerline{\psfig{figure=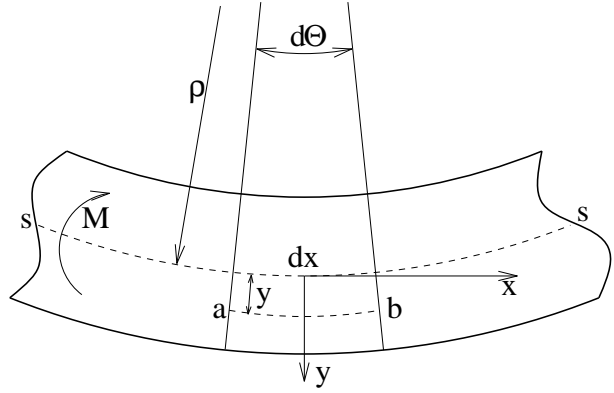,width=8cm,angle=270}}
\vspace{0.5cm}
\fi
\caption{An infinitesimally deformed beam with radius of curvature 
$\rho $. The fibres below or above the neutral fibre $\overline{SS}$ 
are in tension or in compression. For a beam with quadratic cross 
section $b^2$ the deformation leads to the resulting moment 
$M=\frac{E\,b^4}{12\,\rho}$.}
\label{bending.fig}
\end{figure}

When torques and forces are applied to a beam the beam deforms. Since
all deformations are assumed to be small compared with the size of the
beam one can superpose the deformations originating from different
applying forces and
torques~\cite{TimoshenkoGere:1972,TimoshenkoLessells:1928}.
Fig.~\ref{bending.fig} shows an (infinitesimally) deformed beam with
radius of curvature $\rho$. We find the approximation
\begin{equation}
\tan\left( \frac{d\Theta}{2}\right) =\frac{1}{2 \cdot \rho}\cdot
dx \hspace{0.5cm}\mbox{or}\hspace{0.5cm} 
\rho=\frac{dx}{d\Theta}
\end{equation}
with $\tan(d\Theta) = d\Theta$.

The dashed line $\overline{ss}$ is the neutral surface length of which
does not change during the deformation. All other fibres below and
above the neutral surface are in tension or in compression,
respectively. The length of a fibre in distance $y$ from the neutral
surface is $\left( \rho+y\right) d\Theta=\left( 1+y/\rho\right) dx$
and hence the strain reads $\epsilon=y/\rho$. With Hooke's law
$\sigma=E\cdot\epsilon$ one finds $\sigma=Ey/\rho$. Finally the
resulting momentum $M$ is
\begin{equation}
M= \int_A\sigma y dA = \frac{EI}{\rho}
\label{M=EI/rho}
\end{equation}
where the moment of inertia $I=\int y^2 dA$ depends on the geometrical
shape of the cross section of the beam. In the present paper we assume
that the beams have quadratic cross section of width $b$ and hence
$I=b^4/12$. When we express the radius of curvature $\rho$ by
\begin{equation}
\frac{1}{\rho} = \frac{d^2\,v}{dx^2}~,
  \label{curvature}
\end{equation}
where $v$ is the deflection of the beam from its initial position
(with the approximation $\tan{\Theta\approx \Theta}$).
From eqs.~(\ref{M=EI/rho}) and (\ref{curvature}) one finds the basic
equation for the deflection of a beam:
\begin{equation}
\frac{d^2\,v}{dx^2} = v'' = +\frac{M}{E\,I}
  \label{basicequation}
\end{equation}

This equation has to be solved to find the forces and moments which
act when a beam is deformed. The integration constants are used to
satisfy the boundary conditions.

There are three different deformation modes for beams: elongation,
shearing and bending. In the following three subsections
\ref{elongation.subs}, \ref{shearing.subs} and \ref{bending.subs} we
will discuss the deformation rules in detail.

\subsection{Elongation of beams}
\label{elongation.subs}
\begin{figure}[htbp]
  \ifcase \value{nofigs}
\centerline{\psfig{figure=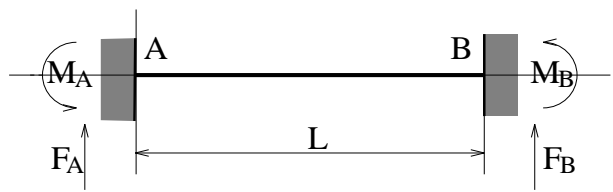,width=8cm,angle=270}}
\vspace{0.5cm}
\fi
    \caption{The shape and the location of a beam at rest. 
Its length is assumed to be $L$, its ends $A$ and $B$ lie 
on the $x$--axis. When the beam is not deformed there do 
not act any forces $F_A$, $F_B$ or moments $M_A$, $M_B$. 
In the following 
\ifcase \value{refcaption}
\ref{bending.fig}--\ref{beamdoppelt.fig} 
\or
figures~8-11
\fi
and in 
\ifcase \value{refcaption}
\ref{shearbend.fig} 
\or
fig.~14
\fi
the $x$--axis is drawn as a thin horizontal line.}

\label{fig:beamatrest}
    
\end{figure}

Since the exact notation of the acting forces and moments as vector
functions of the coordinates of the triangles the beam connects, is
not very instructive but confusing due to the length of the formulae
expressions we discuss for simplicity of notation the deformation of
the beams here and in the following in local coordinates. Instead of
providing the exact vector notation we rather discuss absolute values.
In all cases the directions in which the forces and moments act are
very clear, moreover they are given for each case explicitly in the
figures~\ref{fig:beamatrest}--\ref{beamdoppelt.fig} and in
figure~\ref{shearbend.fig}. For the implementation of the algorithm,
however, one should note that it is necessary to transform the given
expressions for the forces and momenta into the coordinate system of
the triangles. These transformations require a considerable part of
the computation time.

Fig.~\ref{fig:beamatrest} displays the shape and the location of a
beam of length $L$ at rest, i.e.~when no deforming forces $F_A$,
$F_B$ or moments $M_A$, $M_B$ at the ends $A$ and $B$ of the beam
apply. The $x$--axis is drawn with a thin solid line, i.e.~at rest the
beam lies on the $x$--axis.

For the elongation deformation we find according to the
linear Hooke law the restoring force
\begin{equation}
\left| F^{el} \right| = \left| \Delta L \right| \cdot E .
\end{equation}

\subsection{Shearing}
\label{shearing.subs}
\begin{figure}[ht]
  \ifcase \value{nofigs}
\centerline{\psfig{figure=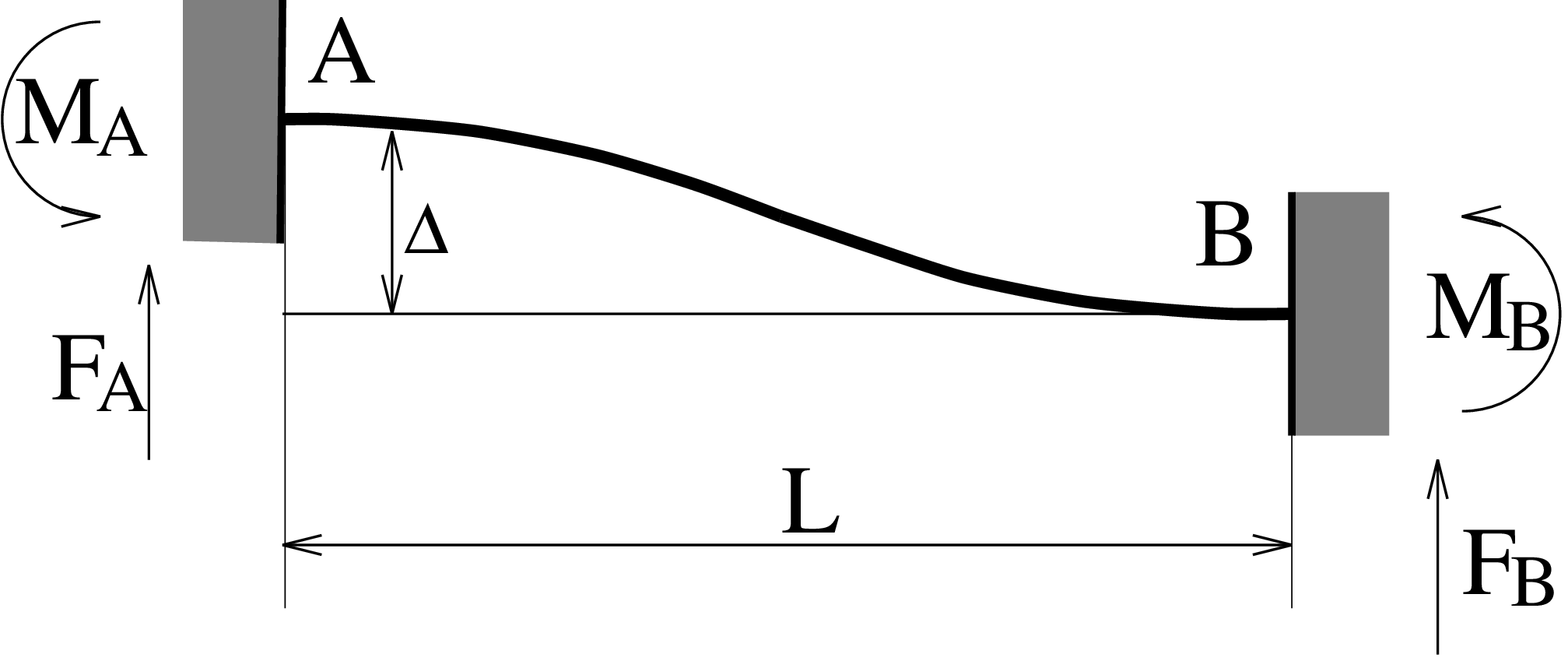,width=8cm,angle=270}}
\vspace{0.5cm}
\fi
\caption{When the beam in 
\ifcase \value{refcaption}
fig.~\ref{fig:beamatrest} undergoes a shear deformation 
$\Delta$ there act the forces $F_A$ and $F_B$ and the moments 
$M_A$ and $M_B$. The forces and moments are given in 
eqs.~(\ref{forcesshearing.eq}, \ref{momshearing.eq}).
%\ref{eq.shearing}).
\or 
fig.~9 undergoes a shear deformation $\Delta$ there act 
the forces $F_A$ and $F_B$ and the moments $M_A$ and $M_B$. 
The forces and moments are given in eqs.~(22).
\fi
}
\label{beamscherung.fig}
\end{figure}

We want to calculate the moments and forces $M_A$, $M_B$, $F_A$ and
$F_B$ of a sheared beam as drawn in Fig.~\ref{beamscherung.fig}.  For
a beam of length $L/2$ which is fixed at one end (at $x=0$) and where
acts the shear force $F$ (in negative $y$--direction) at the other end
(at $x=L/2$) we find for the local moment
$M(x)=F\,\left(\frac{L}{2} - x \right)$ and hence
eq.~(\ref{basicequation}) reads
\begin{equation}
v''=\frac{F}{E\,I} \left(\frac{L}{2}-x\right)
  \label{shearingeq}
\end{equation}
with the boundary conditions $v(0)=0$ and $v'(0)=0$.
We find for the vertical deviation $\Delta/2$ at the point $x=L/2$
\begin{equation}
\Delta/2= \frac{FL^3}{24\,EI}
\end{equation}
force which acts at the free end of the beam.

One can assume that the sheared beam in Fig.~\ref{beamscherung.fig} is
built up of two of such beams of length $L/2$ as described above. 
Finally one finds
\begin{equation}
\Delta = \frac{F\,L^3}{12\,EI}
  \label{delta}
\end{equation}
and hence
%\begin{mathletters}
\begin{eqnarray}
F_A &=& -F_B = F = \frac{12 E I}{L^3}\Delta\label{forcesshearing.eq}\\
M_A &=& M_B = F \cdot \frac{L}{2} = \frac{6 E I}{L^2} \Delta~.
\label{momshearing.eq} 
\end{eqnarray}
%\end{mathletters}

\subsection{Bending}
\label{bending.subs}
\begin{figure}[ht]
  \ifcase \value{nofigs}
\centerline{\psfig{figure=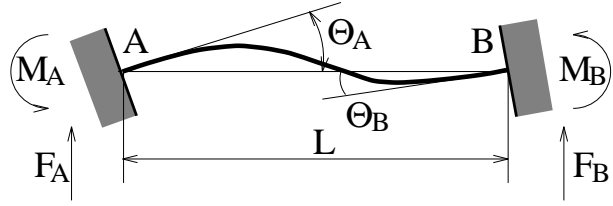,width=8cm,angle=270}}
\vspace{0.5cm}
\fi
\caption{A bent beam. The forces $F_A$ and $F_B$ and the moments $M_A$ 
and $M_B$ can be calculated by superposing the forces and moments of two 
single sided bent beams 
\ifcase \value{refcaption}
(fig.~\ref{beameinseitig.fig}), one of them bent at the side $A$, 
the other one at the side $B$. The results are given in 
eqs.~(\ref{momentsbending.eqA}, \ref{momentsbending.eqB}) and (\ref{forcesbending.eq}).
\or
(fig.~12), one of them bent at the side $A$, the other one at the 
side $B$. The results are given in eqs.~(32) and (33).
\fi
}
\label{beamdoppelt.fig}
\end{figure}

\begin{figure}[ht]
  \ifcase \value{nofigs}
  \centerline{\psfig{figure=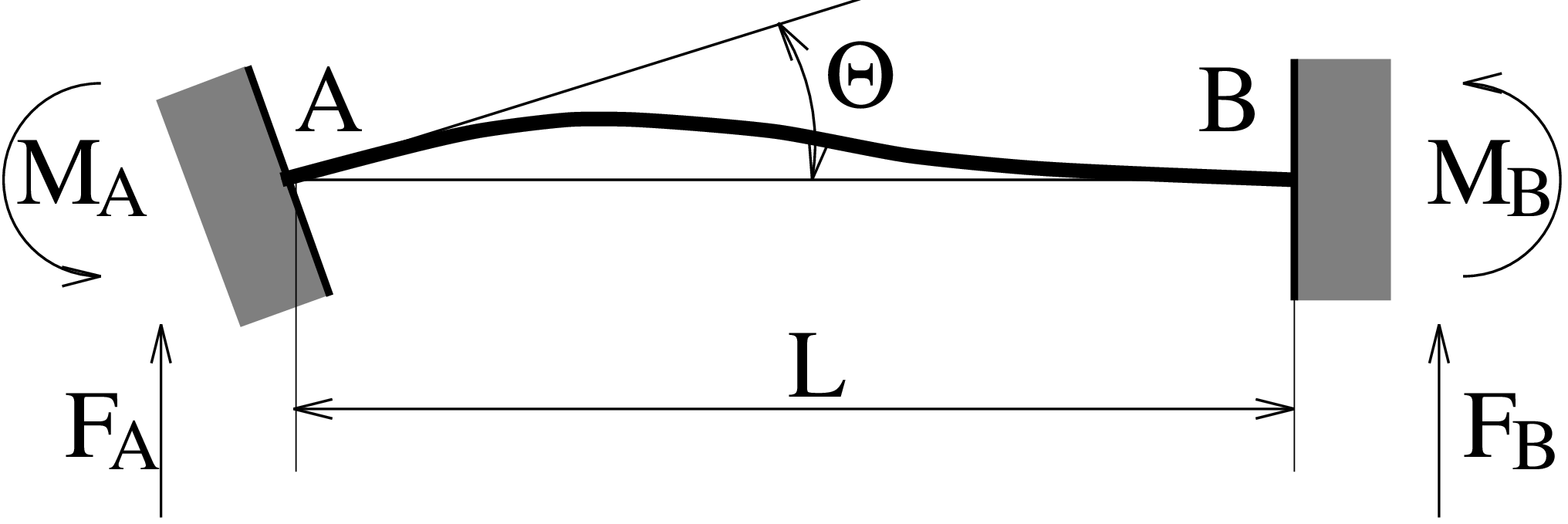,width=8cm,angle=270}}
\vspace{0.5cm}
\fi
\caption{A single sided bent beam. The forces and moments result from the 
superposition of the deformation drawn in 
\ifcase \value{refcaption}
fig.~\ref{auflieger.fig} and the restoring moment $M_B$ 
(fig.~\ref{beameinseitig.fig}) which assures the angle 
$\Theta_B$ to be zero. The equations for the forces and 
moments are given in eqs.~(\ref{onesidebending}) and 
(\ref{onesidebendingf}).
\or
fig.~13 and the restoring moment $M_B$ (fig.~12) which assures the 
angle $\Theta_B$ to be zero. The equations for the forces and 
moments are given in eqs.~(27) and (29).
\fi
}
\label{beameinseitig.fig}
\end{figure}

When a beam undergoes bending deformation
(fig.~\ref{beamdoppelt.fig}), the resulting forces and moments can be
found by superposing the forces and moments that would act if the beam
would be bent at each side separately. Therefore we can restrict
ourself on calculating the forces and moments according to the
single sided deformation drawn in fig.~\ref{beameinseitig.fig}. 

\begin{figure}[ht]
  \ifcase \value{nofigs}
  \centerline{\psfig{figure=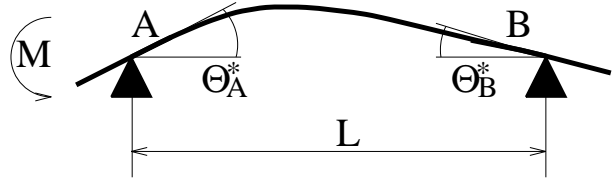,width=8cm,angle=270}}
\vspace{0.5cm}
\fi
\caption{If the ends of the beam are not fixed but freely moving, the 
applied moment $M$ causes bending due to the angles $\Theta_A$ and 
$\Theta_B$ given in 
\ifcase \value{refcaption}
eqs.~(\ref{thetafree.eqA}, \ref{thetafree.eqB}).
\or
eqs.~(23).
\fi
}
\label{auflieger.fig}
\end{figure}

The deformation in fig.~\ref{beameinseitig.fig}, however, can be
understood as a deformation of a beam with free ends according to a
single sided moment $M$ (fig.~\ref{auflieger.fig}) superposed by a
moment $M_b$ (fig.~\ref{beameinseitig.fig}) that assures that the
angle $\Theta_B$ (fig.~\ref{beameinseitig.fig}) results to zero.

First we consider one sided bending of the beam
(fig.~\ref{beameinseitig.fig}). For a beam as drawn in
fig.~\ref{auflieger.fig} where the moment $M$ acts at one of 
its both sides one finds easily
%\begin{mathletters}
\begin{eqnarray}
\Theta_A^* &=& \frac{ML}{3EI}
\label{thetafree.eqA}
\\
\Theta_B^* &=& - \frac{ML}{6EI}.
\label{thetafree.eqB}
\end{eqnarray}
%\end{mathletters}

Since the beam in Fig.~\ref{beameinseitig.fig} has the angle
$\Theta_B=0$ we find the moments acting at the ends of the beam in
fig.~\ref{beameinseitig.fig} by superposing the moment
$M_B$, which causes a virtual angle $\Theta_B^{**}$, fulfilling the
condition $\Theta_B^{**}+\Theta_B^{*}=0$:
\begin{equation}
M_B=\frac{3EI}{L}(- \Theta_B^*)= \frac{M}{2}.
\end{equation} 
The moment $M_B$ causes the angle 
\begin{equation}
\Theta_A^{**}=-\frac{\frac{M}{2}L}{6EI}=-\frac{ML}{12EI}
\end{equation}
and the resulting angle $\Theta_A$ (fig.~\ref{beameinseitig.fig}) is
\begin{equation}
\Theta_A=\Theta_A^{*}+\Theta_A^{**}=\frac{M_AL}{4EI}~.
\end{equation}
Hence, finally we find for the different moments and forces for the bending
deformation (fig.~\ref{beameinseitig.fig})
%\begin{mathletters}
\begin{eqnarray}
M_A &=& - \frac{4EI}{L} \Theta_A \label{onesidebendingA}
\\
M_B &=& - \frac{2EI}{L} \Theta_A \label{onesidebendingB}
\end{eqnarray}
%\end{mathletters}
and with 
\begin{equation}
M_A = - F_BL - M_B
  \label{addmoments}
\end{equation}
\begin{equation}
F_B=-F_A=\frac{6EI}{L^2} \Theta_A~.
\label{onesidebendingf}
\end{equation}
(Note that the moments $M_A$ and $M_B$ in eq.~(\ref{addmoments}) have
to be added as if they would apply to the same end of the beam.)
 
If the beam is bent at both ends by angles $\Theta_A$ and $\Theta_B$
(Fig.~\ref{beamdoppelt.fig}) the resulting moments $M_A$ and $M_B$ and
forces $F_A$ and $F_B$ can be calculated by a superposition of the
forces and moments according to two independent bending deformations
of the type discussed in the previous case.

The angle $\Theta_A$ generates the moments $M_A^{*}$ and $M_B^{*}$
according to eqs.~(\ref{onesidebendingA}, \ref{onesidebendingB})
%\begin{mathletters}
\begin{eqnarray}
M_A^{*} &=& - \frac{4EI}{L} \Theta_A\\
M_B^{*} &=& - \frac{2EI}{L} \Theta_A
\end{eqnarray}
%\label{twosidebendingm}
%\end{mathletters}
and the angle $\Theta_B$ causes the moments
%\begin{mathletters}
\begin{eqnarray}
M_A^{**} &=& - \frac{2EI}{L} \Theta_B \\
M_B^{**} &=& - \frac{4EI}{L} \Theta_B~.
\end{eqnarray}
%\end{mathletters}
Hence we find the resulting moments
%\begin{mathletters}
\begin{eqnarray}
M_A=M_A^{*}+M_A^{**} &=& - \left( \frac{4EI}{L} \Theta_A +
  \frac{2EI}{L} \Theta_B \right)
\label{momentsbending.eqA}
\\
M_B=M_B^{*}+M_B^{**} &=& - \left( \frac{2EI}{L} \Theta_A + \frac{4EI}{L} \Theta_B~\right),
\label{momentsbending.eqB}
\end{eqnarray}
%\end{mathletters}
and with $M_A+M_B=- F_BL$ the resulting forces
\begin{equation}
F_B = -F_A =\frac{6EI}{L^2}\left( \Theta_A + \Theta_B \right)~.
\label{forcesbending.eq}
\end{equation}

Comparing the results from sections \ref{shearing.subs} and
\ref{bending.subs} one can show that each deformation of a beam can be
expressed by bending and elongation, i.e.~the shear deformation need
not to be considered. The proof is given in Appendix \ref{redundanz}.

The deformation of the beams is damped proportionally to the
deformation rates
%\begin{mathletters}
\begin{eqnarray}
  M_A^{(d)} &=& -\frac{\gamma~I}{L} ~\dot{\Theta}_A , \\ 
  M_B^{(d)} &=& -\frac{\gamma~I}{L} ~\dot{\Theta}_B , \\ 
  F_A^{N(d)} &=& -\gamma\left( \vec{v}_A - \vec{v}_B \right)
    \cdot \stackrel{\longrightarrow}{AB} ,\\
  F_B^{N(d)} &=& \gamma\left( \vec{v}_A - \vec{v}_B \right)
    \cdot \stackrel{\longrightarrow}{AB} ,
%  \label{dampingmom}
\end{eqnarray}
%\end{mathletters}
where $\gamma$ is the damping coefficient of the beam material.  As
proven in the appendix the deformation of the beam is in linear
approximation completely determined by the angles $\Theta_A$ and
$\Theta_B$ and the length $L$. Hence there is no damping force acting
in shear direction.

\section{Implementation of the algorithm}
\label{implementationSEC}
We implemented a molecular dynamics algorithm using the particle model
described in the previous sections in FORTRAN. A Gear predictor
corrector scheme of fourth
order~\cite{AllenTildesley:1987,Gear:1966ug} was applied to integrate
Newton's equations of motion. We have to proceed both the forces
caused by the elastic deformation of the triangles as well as the
forces induced by the beams for each iteration step
(eq.~\ref{totalforceEQ}). Since every particle can interact with each
other one causing a high algorithmic complexity, we applied a Verlet
neighborhood list~\cite{Verlet:1967ug} to decrease the amount of
computer time. As discussed below (see table~\ref{table}) the
calculation of the triangle intersections is the most time intensive
part of the calculation. Hence we applied an advanced version of the
neighborhood list method to keep the fraction of the computation of
the triangle--interactions as small as possible.

Our list method acts in two steps. In the first step we prepare
neighborhood lists for each triangle to reduce the number of possible
interactions which has to checked. Because we have to investigate all
possible triangle interactions the time for this step rises as the
square of the total number of triangles $T\sim \left( N \cdot
n_t\right)^2$. (For simplicity we assumed for this estimate that all
$N$ grains consist of $n_t$ triangles each. Hence there are $\left( N
\cdot n_t\right) \cdot \left( \left( N-1\right) \cdot n_t\right) $
possible triangle--interactions.)  In the second step we check for
each entry of the constructed neighborhood list whether the
corresponding triangles interact indeed or not. If they interact,
i.e.~if there is an overlap area, the type of interaction is
determined according to the classification scheme in
section~\ref{trianglesSEC}.

Finally we construct three lists for the interactions of the first
three types of the classification scheme. The interactions of the
forth and fifth type are replaced by interactions of the first type as
described above. The computer time for this step rises linearly with
the number of particles $N$ and with the number of triangles $n_t$
belonging to each particle. Different from usual Verlet tables these
lists contain only triangles which definitely do interact, and hence
we call the most time consuming subroutines for the calculation of the
forces only for triangles which do interact in one of the five manners
described above (section~\ref{trianglesSEC}).

The calculation of the forces induced by the beam deformations follows straight
forward the formulas described in section~\ref{stressesinbeamsSEC}.

In table~\ref{table} we present a detailed performance analysis of
the algorithm. The numbers in the table refer to simulations on a DEC
3000/700 workstation. As visible from the data most of the time
(approximately $70 \%$) is necessary to construct the interaction
list, because we check here every pair of neighbored triangles for
intersections. The construction of this lists give the possibility to
classify the interactions and therefore to use for each kind of
interaction its own optimised algorithm. This prevents us from
calculating useless intersection points, areas and other data.

The computer time needed for the predictor--corrector integration is
very low ($3.7 \%$). Thus it seems to be not useful to optimise the
integration procedure, or to decrease the accuracy of the integration
to safe computer time. The calculation of the beam forces is not
time expensive as well. 

\begin{table}[htbp]
  \caption{The performance analysis of the algorithm. The 
    calculation has been done on a DEC--3000/700 workstation
    (explanation in the text).}
  \begin{tabular}{llll}
    Basic algorithm & portion of & subroutine & portion of\\
    &computation&&computation\\
    \hline
    Construction of the lists & 73.7 \% & Verlet list & 2.07 \% \\
    \cline{3-4}
    && Classifying interaction & 70.62 \% \\ \cline{3-4}
   \hline
    Calculation of the & 7.87 \% & TYPE 1 & 7.54 \% \\ \cline{3-4}
    triangle interaction & & TYPE 2 & 0.24 \% \\ \cline{3-4}
    & & TYPE 3 & 0.09 \% \\
    \hline
    Calculation of the & & & 13.33 \% \\
    beam forces & & &\\
    \hline
    integration & 3.71 \% & predictor & 1.16 \% \\ \cline{3-4}
    && corrector & 2.55 \%\\
    \hline
    all others & & & 1.4 \% 
  \end{tabular}
  \label{table}
\end{table}

From the data in table~\ref{table} it is obvious that triangle
interactions of the first type dominate. Approximately $98 \%$ of the
occurring interactions are of this type. Interactions of the fourth and
fifth type, where our program yields approximative results only, occur
extremely seldomly. We found them only a few times during all our
simulations.

\section{First results}
\label{resultsSEC}
\subsection{The collision of a two particles}
\label{twoparticles}
To demonstrate the non--trivial behavior of colliding non--spherical
particles and to check the correctness of our implemented model we
want to present the results of an experiment where a quadratic grain
collides with another resting one in the absence of gravity. The grains
consist of four equal triangles each, forming a simple quadratic grain
as shown in fig.~\ref{grainsFIG}.  The parameters of the materials
are: $Y=2\cdot 10^{7}\, g/(cm\, sec^2)$, $E = 1\cdot 10^{5}
\,g/(sec^2)$, $I=1\cdot 10^{-4} cm^{3}$ and $\gamma=9$ $g/sec$.
The value of the Young--module $Y$ agrees with most of the simulations
of spheres which can be found in the literature.

\begin{figure}[htbp]
  \ifcase \value{nofigs}
    \centerline{\psfig{figure=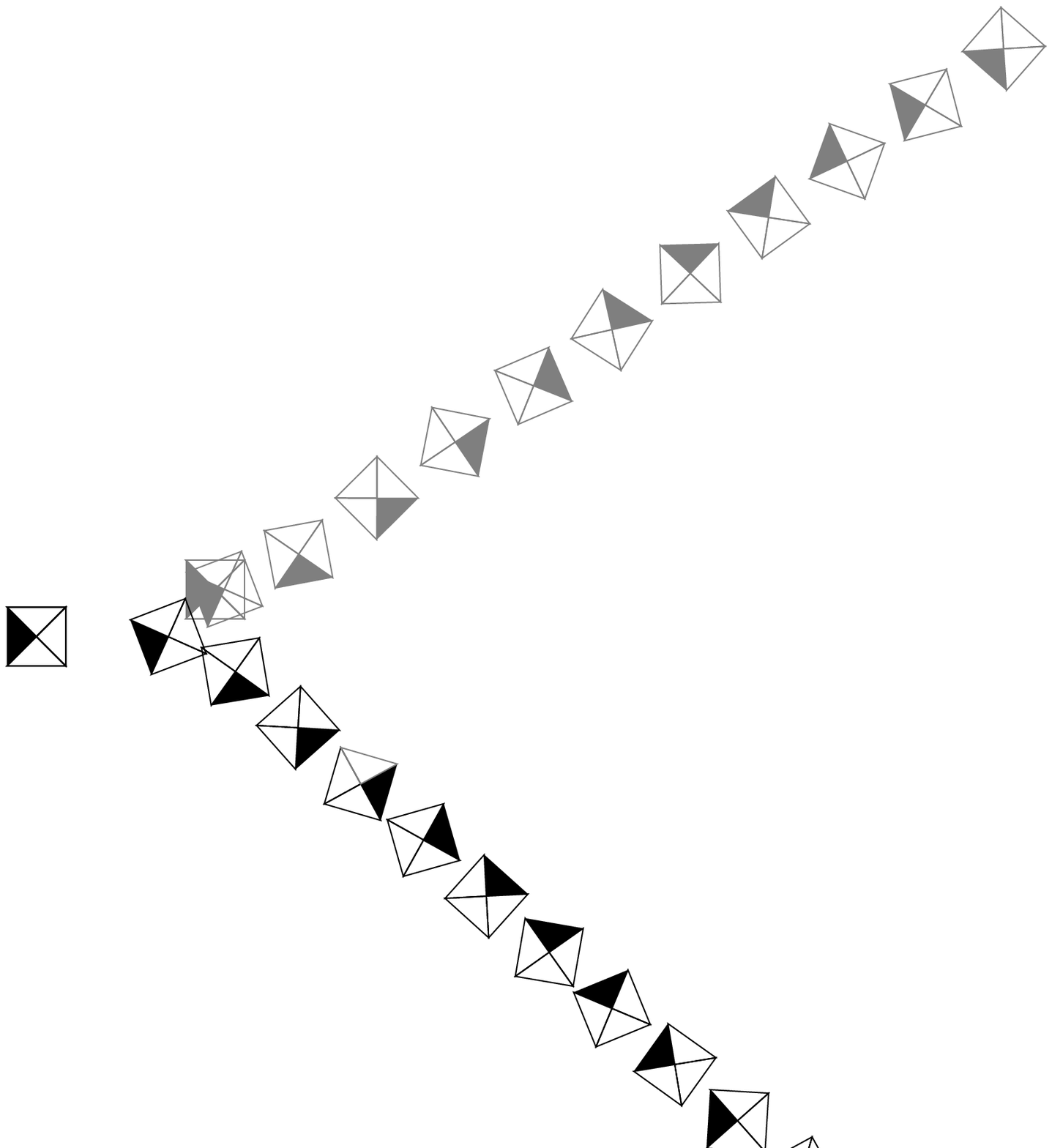,width=4cm}\hspace{0.5cm}\psfig{figure=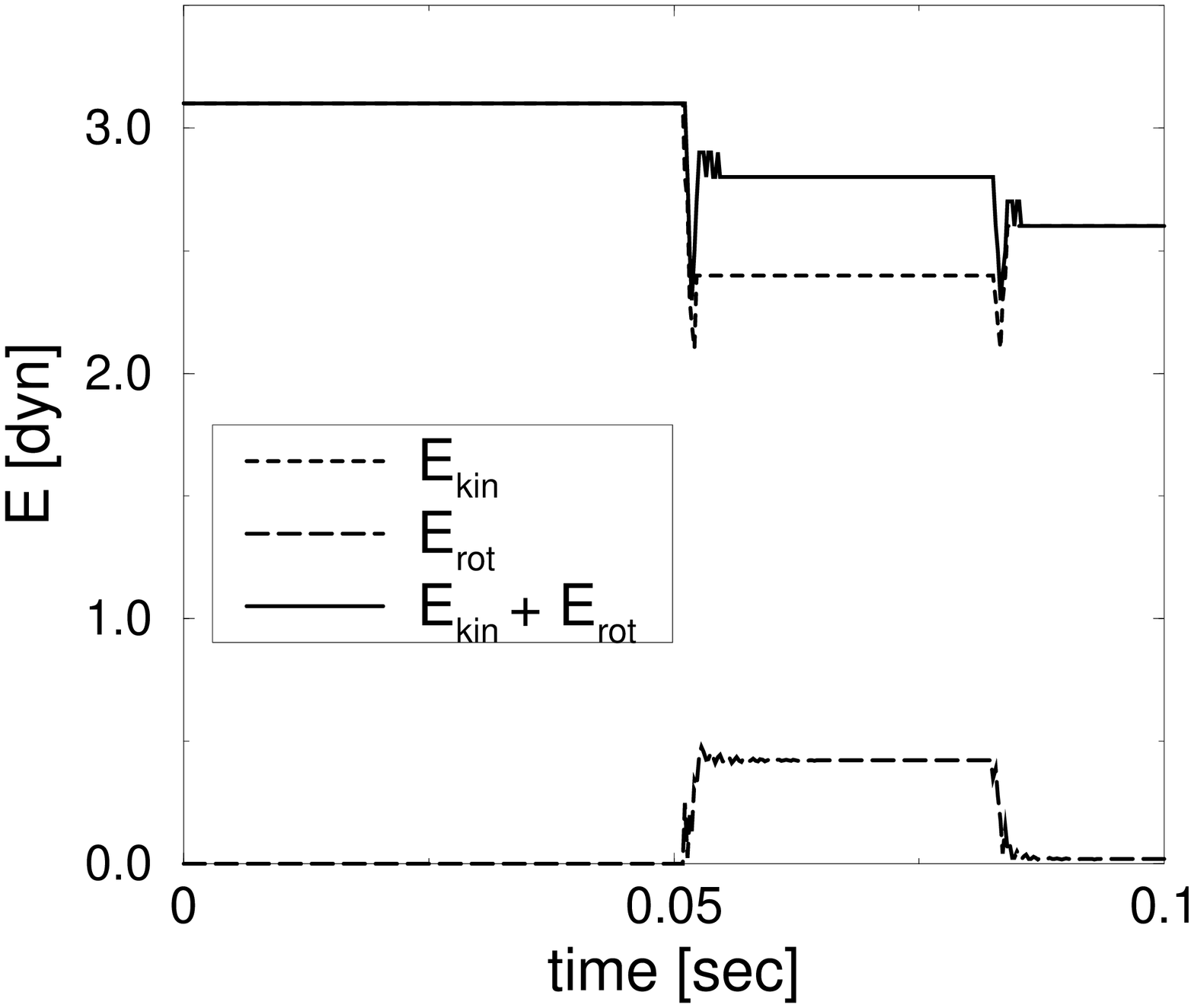,width=7cm,angle=270}}
    \centerline{\psfig{figure=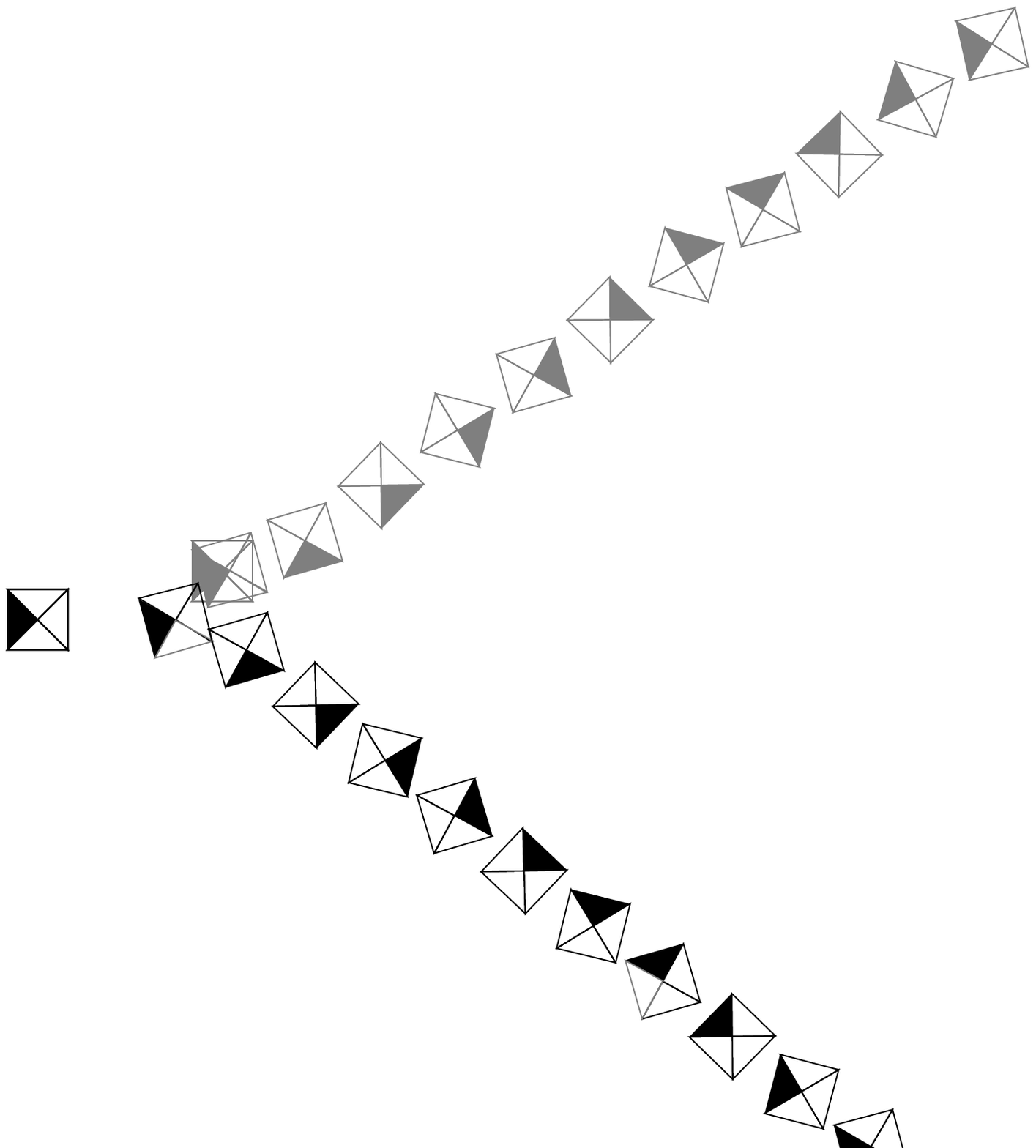,width=4cm}\hspace{0.5cm}\psfig{figure=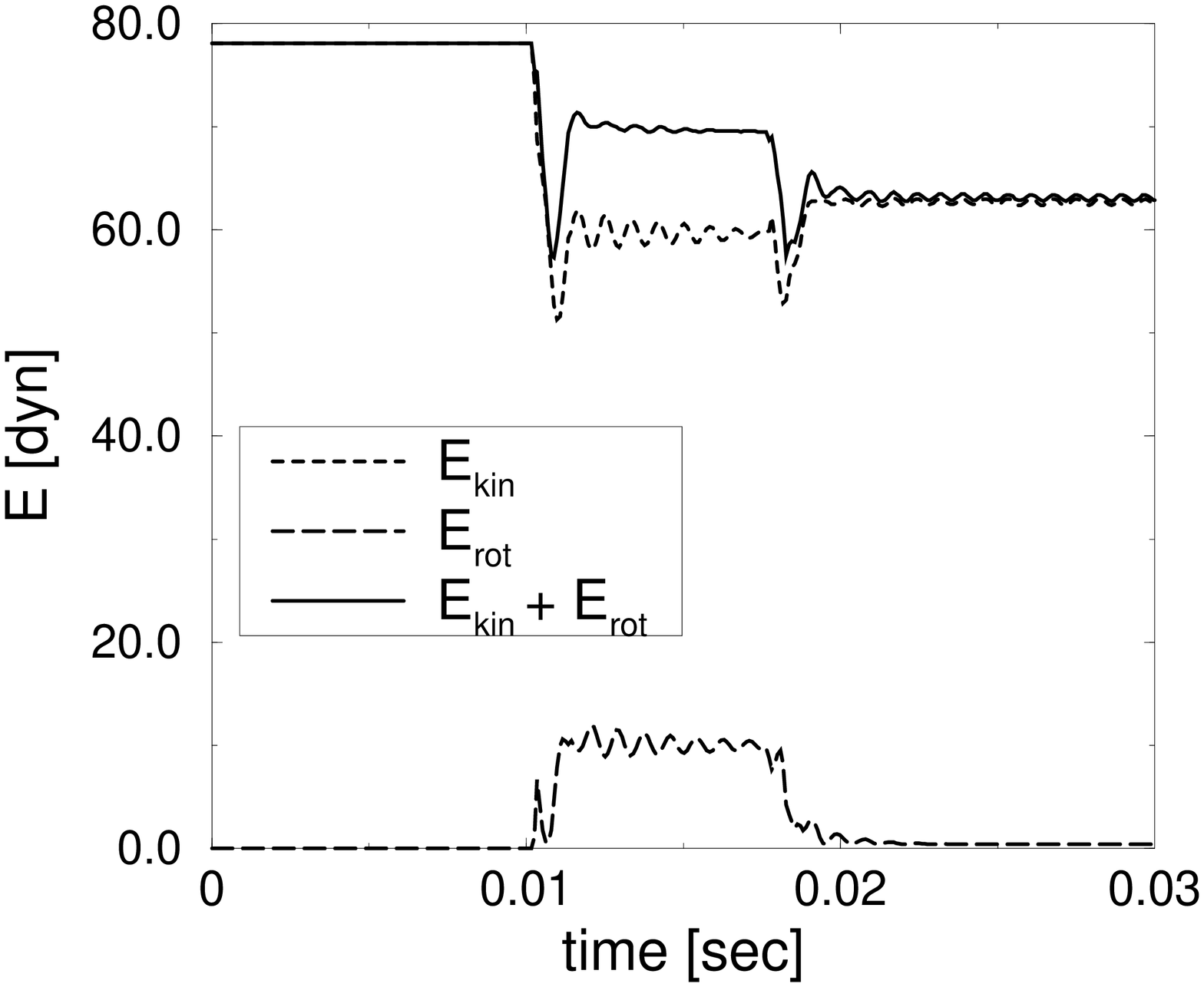,width=7cm,angle=270}}
      \centerline{\psfig{figure=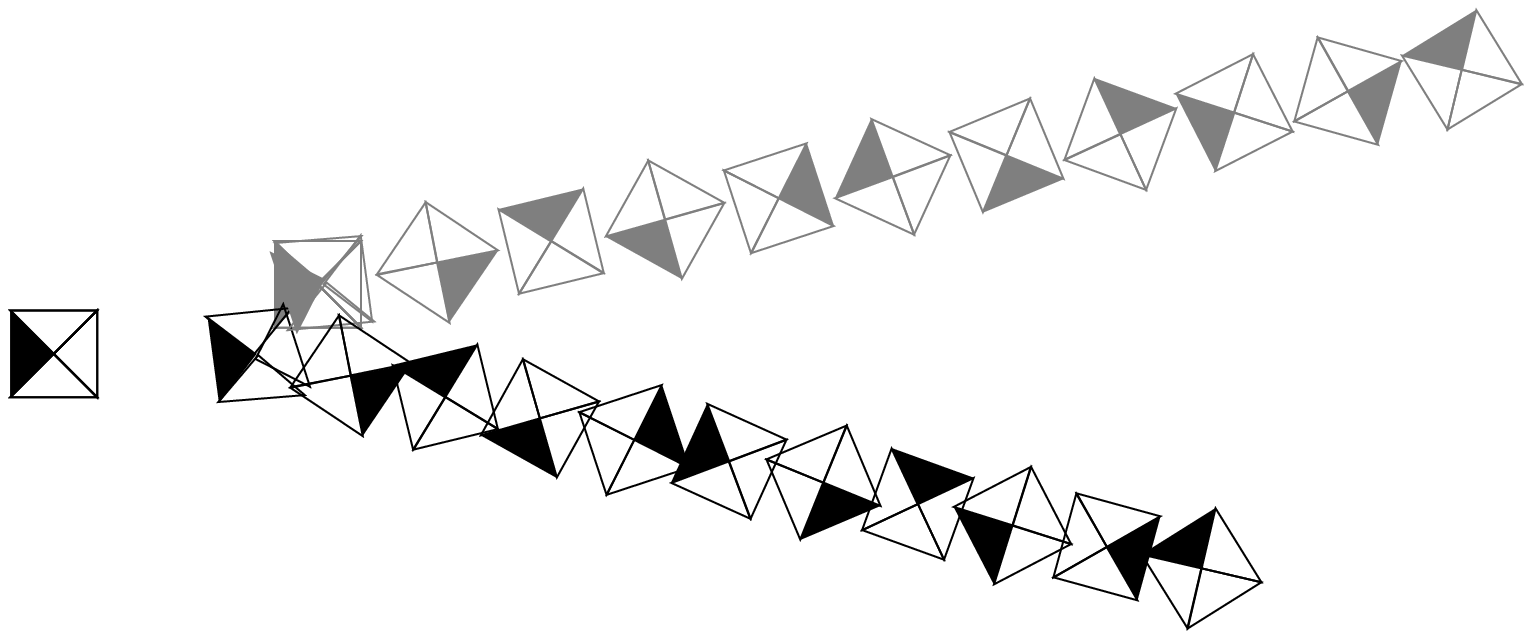,width=4cm}\hspace{0.5cm}\psfig{figure=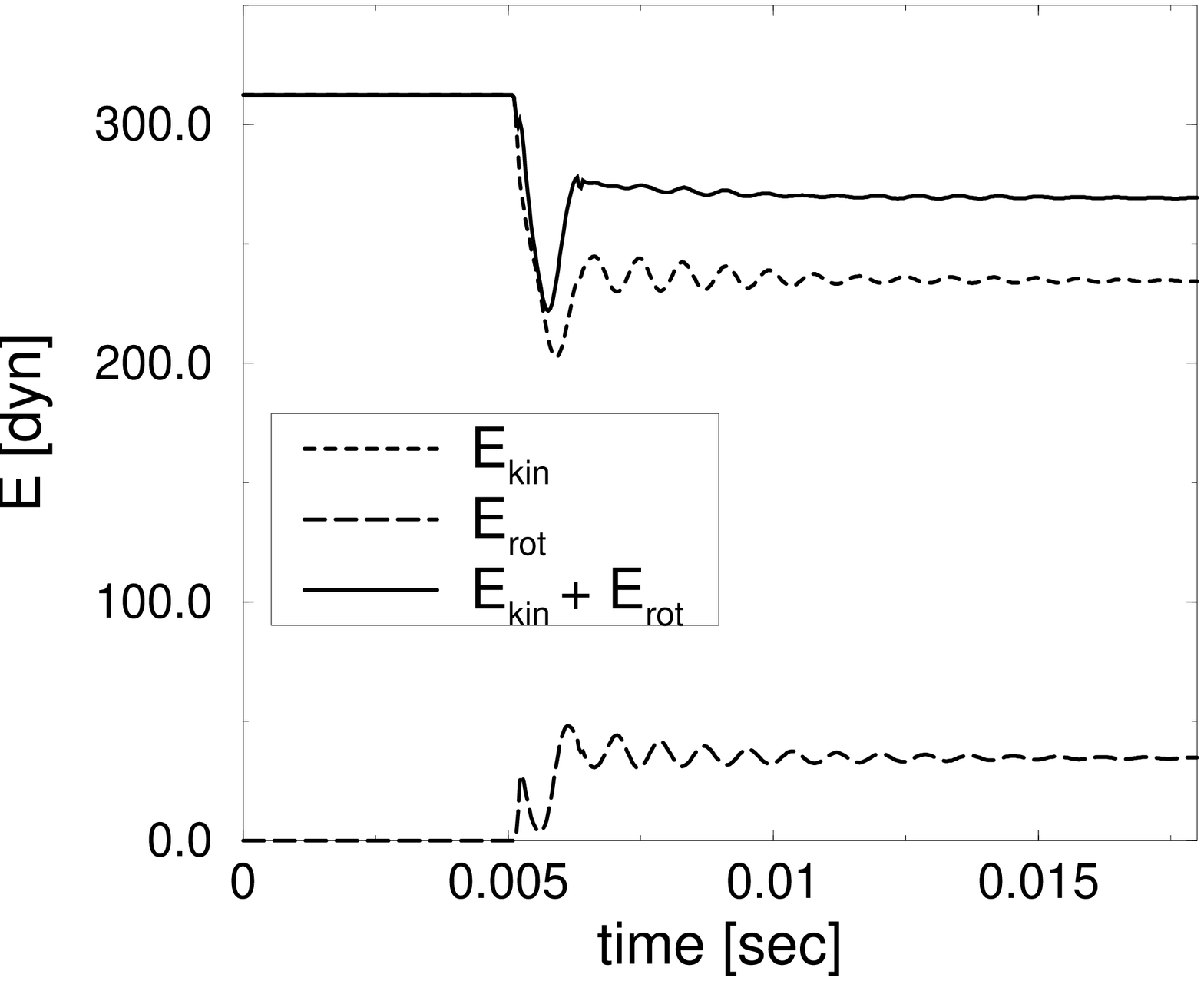,width=7cm,angle=270}}
\vspace{0.5cm}
\fi
  \caption{The collision of a horizontally from left to right 
    moving grain (black drawn) with a resting one (grey) for different
    impact rates $v_{rel}=10 \,cm/sec$ (top), $v_{rel}=50
    \,cm/sec$ (middle), and $v_{rel}=100 \,cm/sec$
    (bottom). The stroboscopic snapshots of the grains are take at
    equidistant time intervals. An animated
    sequence of the collisions can be accessed via World Wide
%???Web~\cite{www}.
Web~[67].
    The right hand figures display the kinetic, rotational and total
    energies of the particle collisions over time.}
  \label{fig:twoparticles}
\end{figure}

Figs.~\ref{fig:twoparticles} (left figures) show the time evolution
of the particles as a sequence of stroboscopic snapshots for different
relative velocities of the particles a)~$v_{rel}=10\,cm/sec$,
b)~$v_{rel}=50\,cm/sec$ and c)~$v_{rel}=100\,cm/sec$.

\begin{figure}[htbp]
  \ifcase \value{nofigs}
  \centerline{\psfig{figure=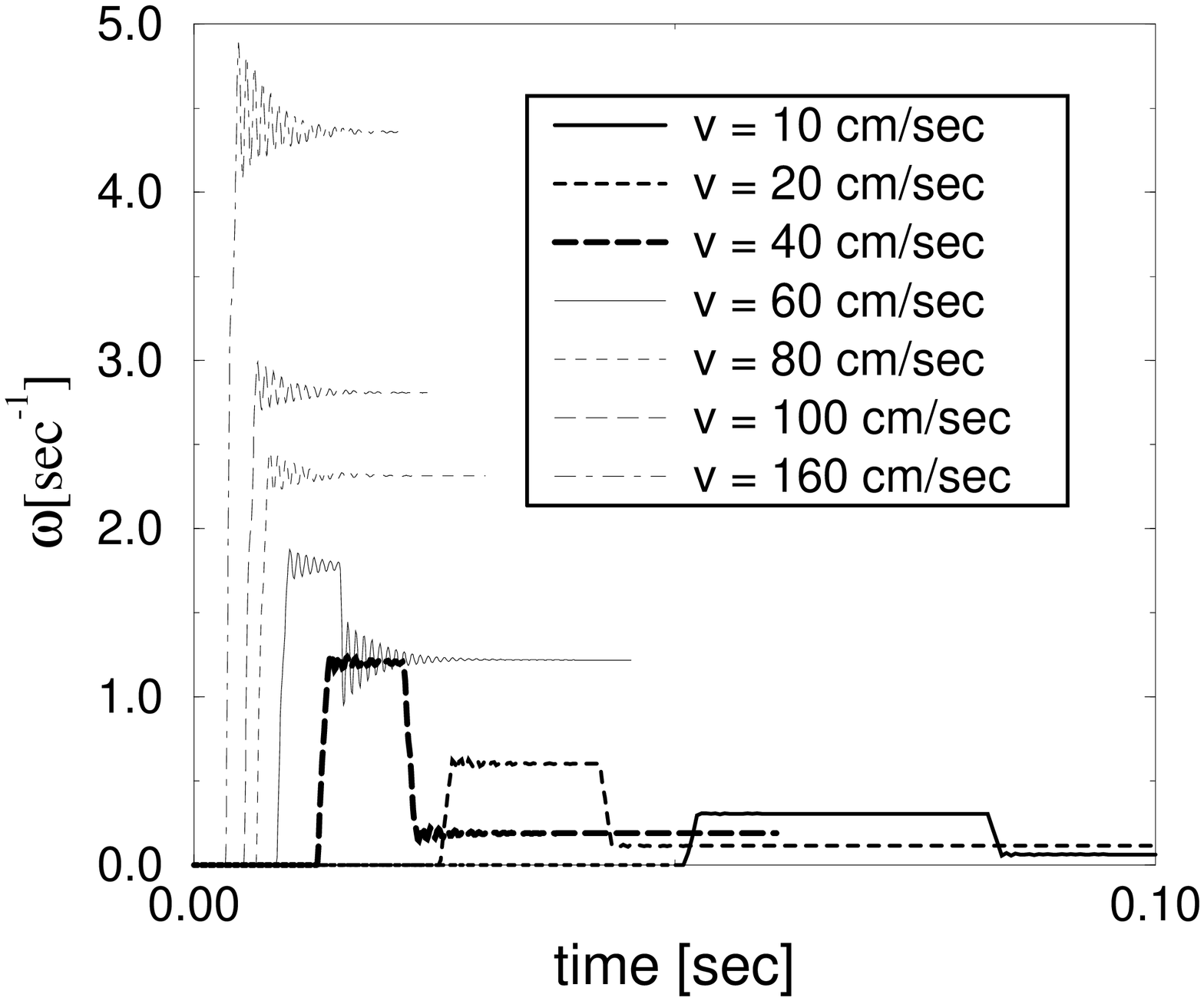,width=8cm,angle=270}}
\vspace{0.5cm}
\fi
  \caption{The angular velocities of the particles over the
    time. For lower velocities the particles collide two times,
    therefore those collisions results in a small angular velocity of
    the grains.}
  \label{rotvel}
\end{figure}

One of the triangles of each grain has been filled to visualise the
rotation of the grains. Initially the velocity of the left particle is
$\vec{v}=\left({v_{rel},0}\right) \,cm/sec$ and the right
particle rests. For low velocities $v<v_{tc}\approx 70 cm/sec$ the
grains collide twice within a very short time. Therefore the traces of
the particles that undergo soft collisions with velocity values
$v<v_{tc}$ differ qualitatively from the traces of particles colliding
with higher relative velocity. Since it is hard to display the
complicated motion of the grains in detail we refer to our World Wide
Web--site URL
http://summa.physik.hu-berlin.de:80/$\sim$thorsten/MDASGP.html~\cite{www}
where one can find animated sequences of various problems mentioned in
the current paper.  The right hand side in fig.~\ref{fig:twoparticles}
shows the rotational and kinetic energies and the total energy as a
function of time. The initial kinetic energy of the translational
motion of the left particle results in kinetic energy of both
particles due to translational motion and rotation. A part of the
mechanical energy is lost due to dissipation in the beams. The
relative amount of rotational energy depends strongly on the impact
velocity as expected from the discussion above. The dependence of the
extinction of the rotational degree of freedom on the impact rate is
demonstrated in fig.~\ref{rotvel} too. This figure shows the time
evolution of the angular velocity of the grains $\omega =
\frac{1}{n_t}\,\left| \sum\limits_{i=1}^{n_t} \left( \vec{v}_i -
    \vec{v}_g \right) \times \vec{s}_i\right| $ over time for
different values of the impact velocity. $\vec{v}_g = \frac{1}{n_t}
\sum\limits_{i=1}^{n_t} \vec{v}_i$ is the velocity of the grain and
$\vec{s}_i$ is the vector pointing from the center of mass point of
the grain to the center of mass point of the triangle $i$. $n_t$ is
the number of triangles the grain consists of. 

For high velocity one
observes an abrupt change of the rotational motion at the time of the
collision and a short damped oscillation according to the excited
inner degrees of freedom of the grain (distortion of beams) and their
relaxation.  After the collision the particles move on with constant
translational and rotational velocities. For low impact rate we find a
quite different behavior: First the particles collide as in the
previous case, resulting in a certain rotational motion. Short time
after the first collision, however, they collide a second time with
different edges. Finally we find for the latter case of slow
collisions a very small resulting angular velocity because the second
collision causes angular moments in opposite directions, hence the
resulting moment is small.

Observing the motion of a moving particle colliding with a resting one
(see~\cite{www}), i.e.~the simplest possible contact of particles, one
can remark that even for this system there is a complex behavior.  For
rigid spheres the velocities after a collision as a function of the
impact velocities~(eq.~(\ref{resitutionEQa}, \ref{resitutionEQb})) can be calculated
analytically solving the viscoelastic equations of the
spheres~\cite{BrilliantovSpahnHertzschPoeschel:1994}. We guess that
this will not be possible for the case of our particles.
      
\subsection{Outflow of a hopper}
\label{sec:hopper}
The outflow of a hopper is of high interest not only because of the
technological importance of this process, but also because of the
exciting phenomena as clogging and density waves which have been
discovered recently (e.g.~\cite{BaxterBehringerFagertJohnson:1989}).
Density waves and clogging have been investigated using molecular
dynamics in two and three dimensions in various papers,
e.g.~\cite{FormKohringMelinPuhlTillemans:1994,LangstonTuezuenHeyes:1993,RistowHerrmann:1995}.
There is at least one other interesting and surprising phenomenon,
detected recently by Evesque and
Meftah~\cite{EvesqueMeftah:1992,EvesqueMeftah:JDP}: They found that a
hour glass ``ticks'' much slower if it is subjected to vertical
vibrations $y=A\cos\left(\omega t\right)$. The effect seems to depend
on the frequency $f=\omega/2\,\pi$ and on the acceleration of the
vibration $\Gamma = A\,\omega^2$. For frequencies between $40\,Hz\le f
\le 60\,Hz$ surprisingly they observed for some accelerations that the
flow almost stops.

We investigated this effect using different particle models and we
found that the effect could be reproduced with the new particle model
only. The results will be described in detail in a forthcoming
paper~\cite{BuchholtzPoeschel:1995hour}.

\begin{figure}[htbp]
  \ifcase \value{nofigs}
\centerline{\psfig{figure=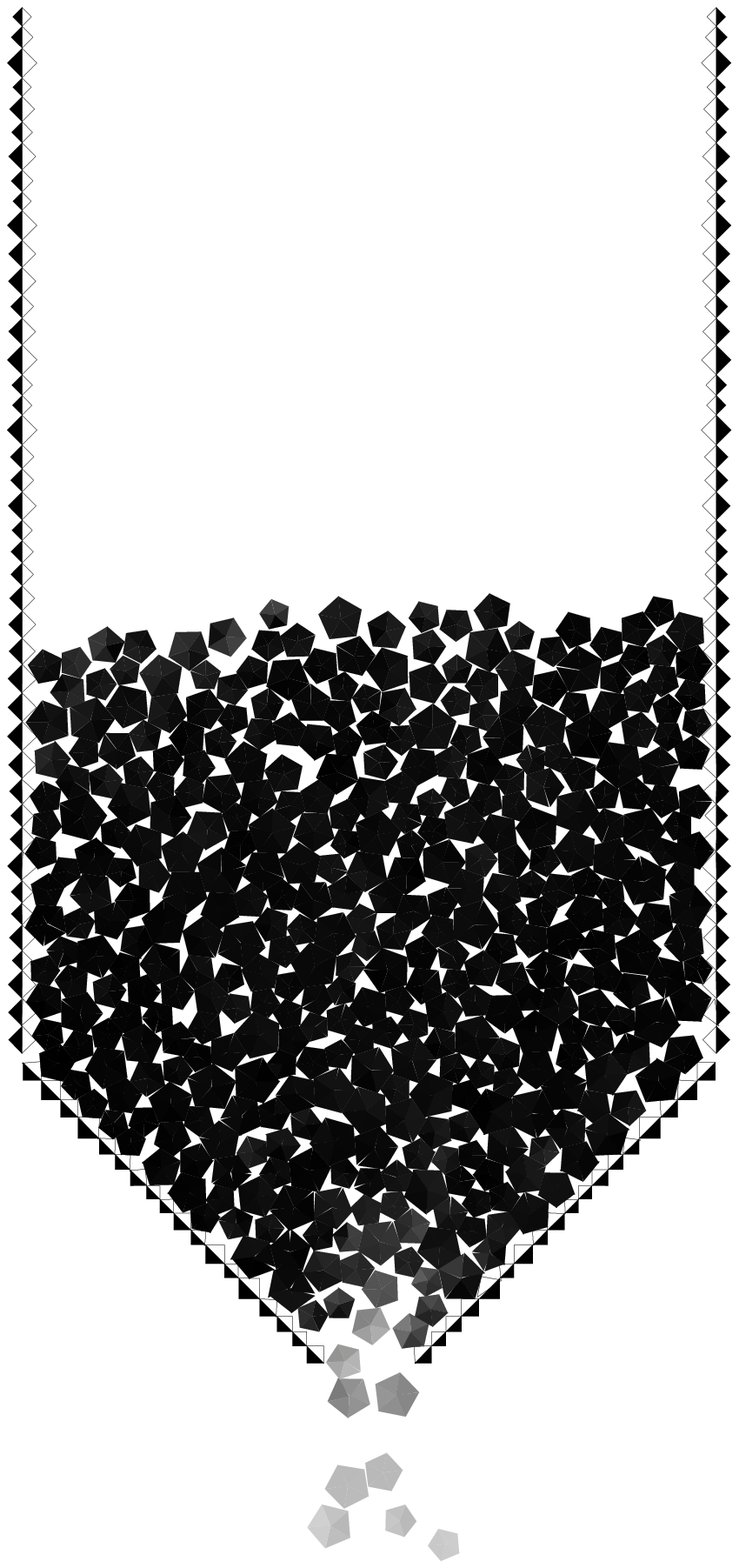,width=4cm,angle=0,bbllx=170bp,bblly=130bp,bburx=430bp,bbury=415bp,clip=}\psfig{figure=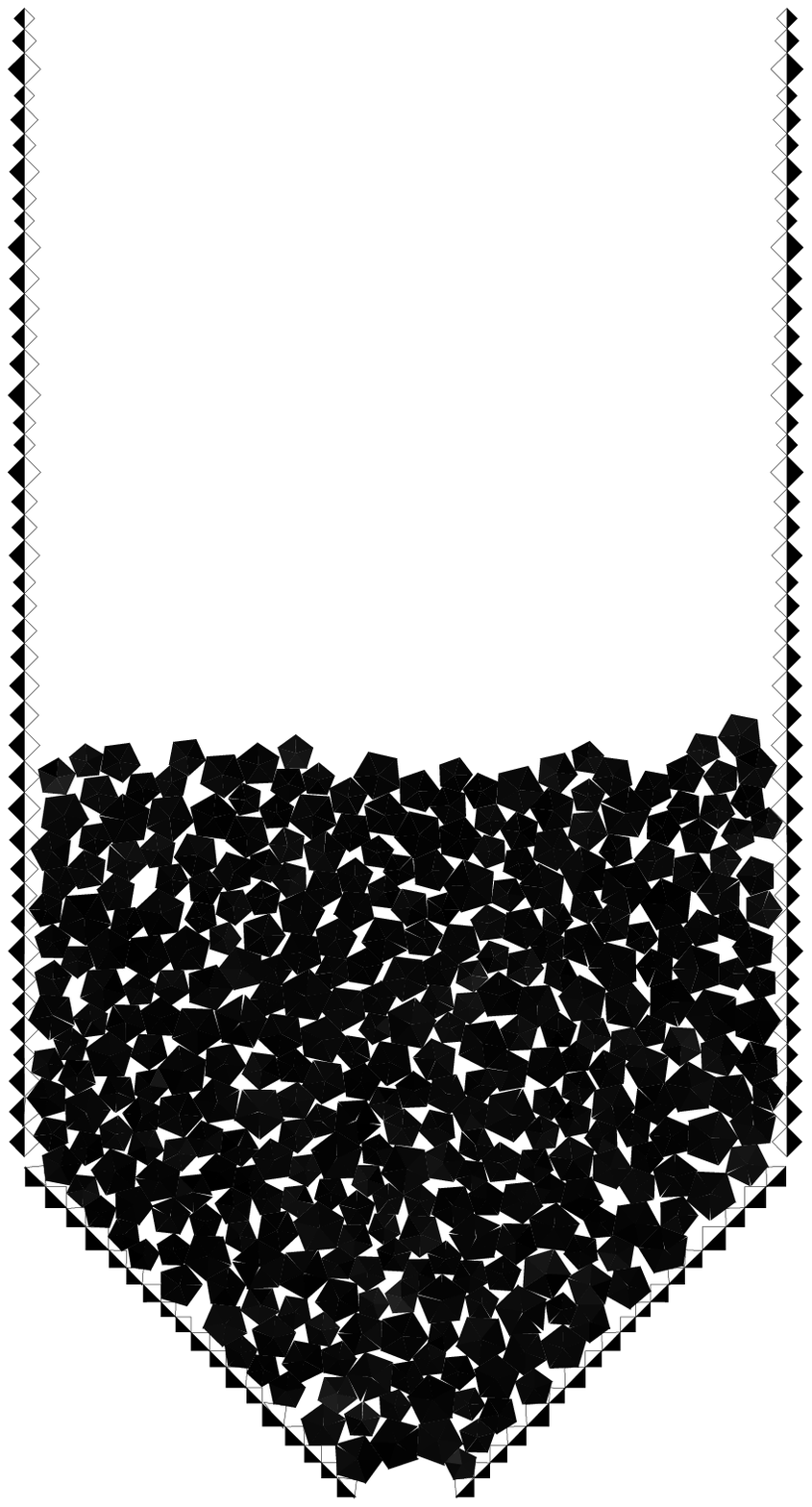,width=4cm,angle=0,bbllx=170bp,bblly=130bp,bburx=430bp,bbury=415bp,clip=}}
\fi
  \caption{Snapshots of the simulation of an outflowing 
    hopper. The left figure shows regular outflow, the right figure
    show a clogging situation. An animated sequence is available via
    World Wide 
%???Web~\cite{www}.
Web~[67].
}
  \label{fig:hopper}
\end{figure}

Fig.~(\ref{fig:hopper}) shows snapshots of the system with $N=400$
complex particles. For an animated sequence we refer to our
WWW--site~\cite{www}. The left figure shows the outflowing hopper, the
flow varies irregularly and the grey scale codes for the particle
velocities. When animating the figures one clearly observes density
waves~\cite{www}. The right hand figure shows the same system a few
moments later. The flow has stopped due to clogging.
%\clearpage

\subsection{Granular flow in a rotating cylinder}
\label{sec:drehtrommel}
The flow of granular material in a rotating cylinder is one of the
most popular problems in the field of granular materials. It has been
investigated experimentally and theoretically by many authors using
various techniques (see
e.g.~\cite{Rajchenbach:1990,Rolf:1993KUG,Nakagawa:1994,ZikEtAl:1994,Ristow:1994,BaumannJanosiWolf:1994,PoeschelBuchholtz:1993}
and many others). The results shall not be discussed here. We applied
the algorithm to this problem too. For the detailed description of the
results see~\cite{BuchholtzPoeschelTillemans:1994}. Here we only want
to demonstrate the abilities of the method and to mention the main
results briefly.

\begin{figure}[ht]
  \ifcase \value{nofigs}
\centerline{\psfig{figure=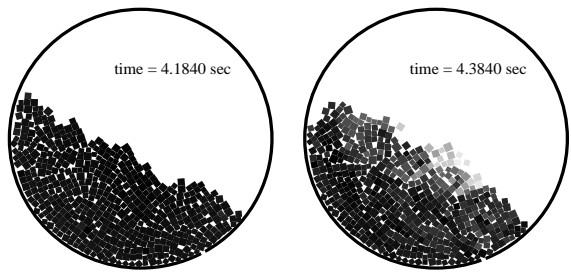,width=8cm,angle=0}}
\fi
\caption{\em Snapshots of the simulation with $N=500$ complex
  particles with $P\in [0.1\,cm; 0.2\,cm]$ in a rotating cylinder of
  diameter $D=4\,cm$. The angular velocity is $\Omega=0.1\,sec^{-1}$.
  The wall particles have not been drawn. The right snapshot shows an
  avalanche, the left one has been taken short time before. The grey
  scale codes for the particle velocity.}
\label{snap.fig}
\end{figure}

Fig.~\ref{snap.fig} shows snapshots of the simulation of a slowly
rotating cylinder. The grey scale codes for the velocity of the
grains, black means high velocity, black codes for high velocity.
For low angular velocity of the driven cylinder one finds
experimentally stick--slip flow (e.g.~\cite{Rajchenbach:1990}),
i.e.~the material moves downwards not homogeneously but it forms
avalanches. In the right figure one observes an avalanche on the top
of the material indexed by light grey shadowed grains. The left figure shows
the same system immediately before the avalanche. Avalanches are
relatively seldom events. Most of the time the particles rest with
respect to each other, i.e.~they move only due to the rotation of the
cylinder (left side of fig.~\ref{snap.fig}). When increasing the
angular velocity of the cylinder there is a relatively sharp
transition between the stick--slip flow and the homogeneous regime.
This transition could be found in the simulation using non--spherical
grains~\cite{BuchholtzPoeschelTillemans:1994}. Animated sequences for
both regimes, stick--slip and continuous flow, can be accessed via
World Wide Web~\cite{www}.

\begin{figure}[ht]
  \ifcase \value{nofigs}
\centerline{\psfig{figure=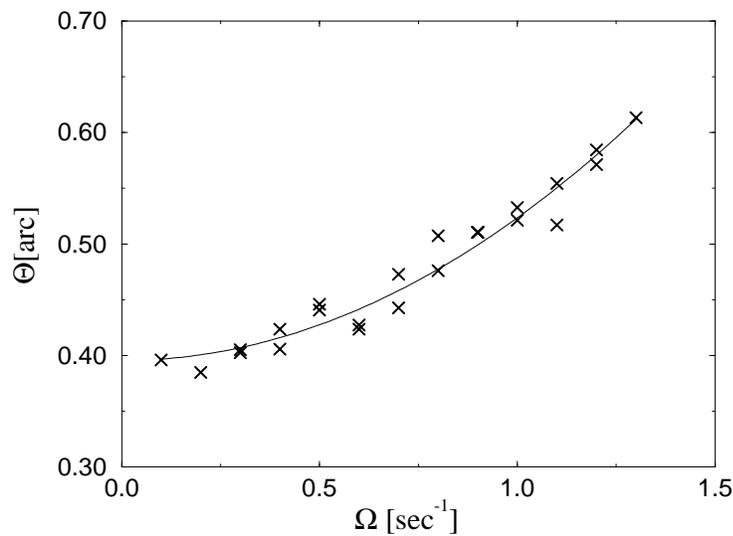,width=10cm,angle=270}}
\vspace{0.5cm}
\fi
\caption{\em The inclination $\Theta$ of the material surface over 
  the angular velocity $\Omega$. The dotted line displays the function
  which has been measured experimentally by
%???  \nobreak{Rajchenbach}~\cite{Rajchenbach:1990}.
  \nobreak{Rajchenbach}~[73].
}
\label{inclination.fig}
\end{figure}

Fig.~\ref{inclination.fig} shows the dependence of the difference
between the inclination of the material and the angle of repose on the
angular velocity of the cylinder. The line displays the behavior found
by Rajchenbach $\Theta-\Theta_c \sim
\Omega^2$~\cite{Rajchenbach:1990}. The simulation data are in good
agreement with this observation. 

Both results mentioned, the sharp transition between the flow regimes
and the inclination as a function of the angular velocity of the
cylinder, agree with experimental observations. We have not been
able to find these effects using simple spheres or using particles
composed of
spheres~\cite{PoeschelBuchholtz:1993,PoeschelBuchholtz:1993CSF} in
molecular dynamics simulations.

\section{Conclusion}
\label{ConclusionSEC}
We presented a model for the simulation of the particles of a granular
material. Each particle consists of triangles which are connected by
deformable beams. There are no restrictions concerning the shape of the
grains (convex or concave), nor the number or the shape of the
triangles a grain consists of. To preserve calculation time it is
favourable to chose arrangements of the triangles so that the beams
are fixed at the center of mass points of the triangles. Otherwise one needs
additional computation time caused by the Steiner law. 

When two triangles of different grains collide there acts an elastic
restoring force proportionally to the compression of the particles.
During the collision of the triangles the energy is preserved. A
collision of a triangle with another one causes moments and forces
acting on this triangle and hence a deformation of the beams which
connect the triangles to neighboring ones. The deformation of the
beams causes forces and moments acting on the neighboring triangles.

Beams can be deformed in axial direction (elongation) and in shear
direction and they can be bent. When a beam is deformed it dissipates
energy proportionally to the deformation rate.

In the present paper we described beams that recover completely while
dissipating mechanical energy when the deforming moments and forces
vanish. Our model, however, is not restricted to this case.
Interesting phenomena are plastic deformation and wear which can be
easily simulated by introducing thresholds for the beam forces and
moments or for the deformations. When a force or a moment or a
deformation, respectively, exceeds the threshold the beam breaks or
deforms permanently. Similar simulations with other models have been
reported in the
literature~\cite{Tillemans:1994,Handley:1993,PotapovCampbellHopkins:1994aug}.

The proposed algorithm has been implemented in FORTRAN. We have
demonstrated the behavior of a system of grains applying it to three
examples of granular assemblies. In a simple collision simulation of
two grains we discussed the trajectories of the grains as well as the
evolution of the kinetic and rotational energies depending on the
impact rate. Sample results for the flow out of a hopper and the
motion of granular material in a rotating cylinder have been
presented. In the latter case we found quantitative agreement of the
simulation with the experiment that could not be found so far, neither
with spherical grains nor with non--spherical grains of other type.

\ack We thank Hans J.~Herrmann, Stefan Schwarzer and
Hans--J\"urgen Tillemans for many helpful discussions and comments.
This work has been supported by the Deutsche Forschungsgemeinschaft
(grant Ro 548/5-1).

\begin{appendix}
\section{Shear deformation can be expressed by bending}
\label{redundanz}
The deformation of a beam in the two dimensional space has three
degrees of freedom: the distance of the relative positions of the ends
of the beams $L=\sqrt{\left( x_A-x_B\right) ^2+\left( y_A-y_B\right)
  ^2}$, and the angles $\Theta_A$ and $\Theta_B$. These values are the
parameters in eqs.~(\ref{momentsbending.eqA}, \ref{momentsbending.eqB}) and
(\ref{forcesbending.eq}). The shear parameter $\Delta$ in
eqs.~(\ref{forcesshearing.eq}, \ref{momshearing.eq})
%\ref{eq.shearing}) 
is a fourth
parameter, i.e.~one of them must be redundant. In the following we
will show that shear deformation can be expressed by bending, at least
in our approach of a linear deformation--force relation.

\begin{figure}[ht]
  \ifcase \value{nofigs}
\centerline{\psfig{figure=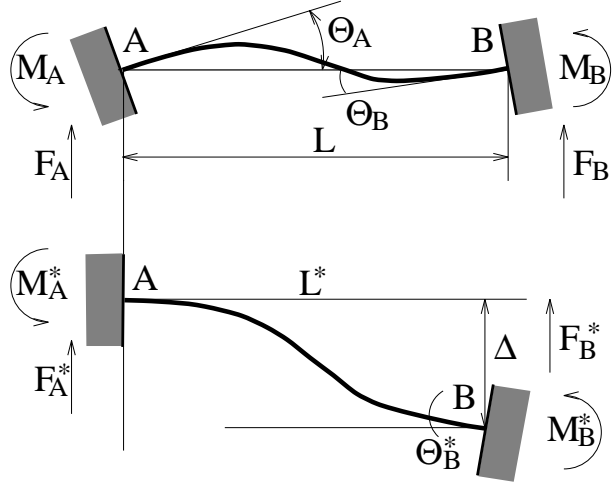,width=8cm,angle=270}}
\vspace{0.5cm}
\fi
\caption{The deformation of a beam in shear direction $\Delta$ (upper 
part) can be represented by bending deformation (lower part). The 
appendix contains the proof that one need not to consider shear 
{\em and} bending deformation in the linear approximation which 
is justified for small deformations.}
\label{shearbend.fig}
\end{figure}
Suppose we have a deformed beam as drawn in Fig.~\ref{shearbend.fig}
(bottom). Then applying eqs.~(\ref{forcesshearing.eq}, \ref{momshearing.eq})
%\ref{eq.shearing}) 
we find for the
forces and moments according to shearing
%\begin{mathletters}
\begin{eqnarray}
M_A^s &=& M_B^s = \frac{6EI}{\left(L^*\right)^2}\, \Delta\\
F_A^s &=& -F_B^s = \frac{12EI}{\left(L^*\right)^3}\, \Delta
%\label{Shearing*}
\end{eqnarray}
%\end{mathletters}
and according to bending (eqs.~(\ref{momentsbending.eqA},\ref{momentsbending.eqB}) and
(\ref{forcesbending.eq}))
%\begin{mathletters}
\begin{eqnarray}
M_A^b &=& -\frac{2EI}{L^*}\,\Theta_B^*\\
M_B^b &=& -\frac{4EI}{L^*}\,\Theta_B^*\\
F_A^b &=& -F_B^b = -\frac{6EI}{\left(L^*\right)^2}\,\Theta_B^*~.
%\label{Bending*}
\end{eqnarray}
%\end{mathletters}
The total forces and moments read
%\begin{mathletters}
\begin{eqnarray}
M_A^*&=& M_A^s + M_A^b = \frac{6EI}{\left(L^*\right)^2}\,\Delta - \frac{2EI}{L^*}\,\Theta_B^* 
\label{Total*A}
\\
M_B^*&=& M_B^s + M_B^b = \frac{6EI}{\left(L^*\right)^2}\,\Delta - \frac{4EI}{L^*}\,\Theta_B^*
\label{Total*B}
 \\
F_A^*&=&-F_B^* = \frac{12EI}{\left(L^*\right)^3}\, \Delta - \frac{6EI}{\left(L^*\right)^2}\,\Theta_B^*~.
\label{Total*C}
\end{eqnarray}
%\end{mathletters}
Turning the entire system by the angle $\Theta_A\approx\tan \Theta_A =
-\frac{\Delta}{L^*}$ we obtain the new system as drawn in the top of
fig.~\ref{shearbend.fig} with the new angles
$\Theta_A=-\frac{\Delta}{L}$ and $\Theta_B=\Theta_B^*-\frac{\Delta}{L}$.
With eqs.~(\ref{momentsbending.eqA}, \ref{momentsbending.eqB}) and (\ref{forcesbending.eq}) we find
%\begin{mathletters}
\begin{eqnarray}
M_A &=& -\frac{4EI}{L}\,\Theta_A- \frac{2EI}{L}\,\Theta_B = \frac{6EI}{L^2}\,\Delta - \frac{2EI}{L}\,\Theta_B^*
\label{TotalA}
\\
M_B &=& -\frac{2EI}{L}\,\Theta_A- \frac{4EI}{L}\,\Theta_B = \frac{6EI}{L^2}\,\Delta - \frac{4EI}{L}\,\Theta_B^*
\label{TotalB}
\\
F_A &=& -F_B = -\frac{6EI}{L^2}\,\left(\Theta_A+\Theta_B \right) = \frac{12EI}{L^3}\,\Delta - \frac{6EI}{L^2}\,\Theta_B^*
\label{TotalC}
\end{eqnarray}
%\end{mathletters}
Approximating $L$ by $\sqrt{L^2+\Delta^2} = L^*$ evidently the results
in eqs.~(\ref{TotalA}-\ref{TotalC}) coincide with eqs.~(\ref{Total*A}-\ref{Total*C}). Thus we do
not need to consider shear deformation in our simulations since it is
redundant here.
\end{appendix}

\end{document}